\documentclass[twocolumn,prd,amsmath,amssymb,epsf]{revtex4-1}

\usepackage[dvipdfmx]{graphicx}
\usepackage{dcolumn}
\usepackage{bm}
\usepackage{mathrsfs}

\begin{document}


\title{Escape of cosmic rays from perpendicular shocks in the circumstellar magnetic field}
\author{Shoma F. Kamijima}
\author{Yutaka Ohira}
\affiliation{Department of Earth and Planetary Science, The University of Tokyo, 7-3-1 Hongo, Bunkyo-ku, Tokyo 113-0033, Japan}


\begin{abstract}
We investigate the escape process of cosmic rays (CRs) from perpendicular shock regions of a spherical shock propagating to a circumstellar medium with the Parker-spiral magnetic field.
The diffusive shock acceleration in perpendicular shocks of supernova remnants (SNRs) is expected to accelerate CRs up to PeV without upstream magnetic field amplification.
Red supergiants (RSGs) and Wolf-Rayet (WR) stars are considered as progenitors in this work.
We perform test particle simulations to investigate the escape process and escape-limited maximum energy without magnetic field amplification in the upstream region, where the magnetic field strength and rotation period expected from observations of RSGs and WR stars are used. 
We show that particles escape to the far upstream region while moving along the equator or poles and the maximum energy is about $10-100~{\rm TeV}$ when SNRs propagate to free wind regions of RSGs and WR stars.
In most cases, the escape-limited maximum energy is given by the potential difference between the equator and pole.
If progenitors are oblique rotators and SNRs are in the early phase just after the supernova explosion, the escape-limited maximum energy is limited by the half wavelength of the wavy current sheet.
In addition, for RSGs, we show that the luminosity of CRs accelerated in the wind region is sufficient to supply the observed CR flux above $10~{\rm TeV}$ if a strong magnetic field strength is sustained in most RSGs.
In terms of the CR luminosity, SNRs propagating to the free wind of WR stars can contribute to PeV CRs.
As long as no magnetic field amplification works around SNR shocks, the maximum energy is decided by the magnetic field strength in the wind region, which depends on the rotation period, stellar wind, and surface magnetic field of RSGs and WR stars.
Therefore, we need to observe these quantities to understand the origin of CRs.
\end{abstract}

\maketitle

\section{Introduction} \label{sec:intro}
The origin of cosmic rays (CRs) is a longstanding problem since the discovery of CRs.
It is believed that supernova remnants (SNRs) are accelerator of Galactic CRs below $3~{\rm PeV}$.
The HESS experiment observed a PeVatron candidate around the Galactic center \cite{hess}.
HAWC found some PeVatron candidates in our Galaxy \cite{hawc}.
Furthermore, Tibet AS$\gamma$ reported gamma rays above $100~{\rm TeV}$ from the potential PeVatron (G106.3+2.7) and sub-PeV diffuse gamma rays from the Galactic disk \cite{tibet}.
LHAASO observed gamma rays above $100~{\rm TeV}$ from 12 Galactic sources \cite{lhaaso}.
In addition to the above experiments, it is expected that many PeVatron candidates will be found by ALPACA \cite{alpaca} and SWGO \cite{swgo} that locate in the southern hemisphere.
In terms of energy spectra of CR protons and heliums, CREAM, NUCLEON, DAMPE, and HAWC experiments reported the spectral break around $10~{\rm TeV}$ \cite{10tev}. 
The PeV scale is thought to be the maximum energy scale of Galactic CR protons, but it is still unknown what the energy scale around $10~{\rm TeV}$ means.

The diffusive shock acceleration (DSA) is the plausible acceleration mechanism that  accelerates CRs up to PeV \cite{cr}.
In DSA, particles cross the shock front many times while diffusing in both upstream and downstream regions, gaining energy by numerous shock compressions.
The acceleration time of the DSA depends on the angle between the shock normal direction and magnetic field \cite{drury83}.
Therefore, the standard picture for CR acceleration up to PeV is different between parallel shocks and perpendicular shocks, in which the shock normal direction is parallel and perpendicular to the magnetic field direction, respectively.
For the DSA at parallel shocks, upstream and downstream magnetic fields have to be on the order of $100~{\rm \mu G}$ to make the maximum energy the PeV scale \cite{cesarsky81}. 
Since the typical magnetic field strength is about $3~{\rm \mu G}$ in the interstellar medium (ISM), the magnetic field has to be strongly amplified.
Many authors proposed magnetic field amplification  mechanisms \cite{bell04, crsi}, but the amplification is still controversial. 
Even without the magnetic field amplification, CRs could be accelerated to the PeV scale by the adiabatic compression of pulsar wind nebulae inside SNRs \cite{ohira18}. 
Recently, it was proposed that CRs are accelerated to the PeV scale by a shock propagating to a dense wind in two weeks after supernova explosions \cite{inoue21}.
Type IIn supernovae are thought to occur in this dense wind region \cite{smith14}.

On the other hand, perpendicular shocks are thought to accelerate CRs to the PeV scale without upstream magnetic field amplification \cite{jokipii87}.
This is because acceleration at the perpendicular shock is faster than at the parallel shock  \cite{takamoto15, kamijima20}.
The rapid acceleration at the perpendicular shock was confirmed by numerical simulations \cite{rapidperp}. 
As for the momentum spectrum of accelerated particles at the perpendicular shock, it was claimed that the momentum spectrum is softer than that of the standard DSA, $dN/dp \propto p^{-2}$, when the magnetic field fluctuation is weaker than the uniform magnetic field component in both upstream and downstream regions to realize the rapid acceleration \cite{takamoto15}.
However, recent observations and numerical simulations suggest that the magnetic field fluctuation in the downstream region is highly turbulent \cite{berezhko03, bamp}. 
In this case, the momentum spectrum is the canonical spectrum, $dN/dp \propto p^{-2}$, even at the perpendicular shock \cite{kamijima20}.

The maximum energy of CRs is limited not only by the age of SNRs but also by escape from SNRs \cite{ptuskin03}. 
The escape of CRs plays an important role to determine the energy spectrum of observed CRs \cite{ohira10}.
The shape of the shock surface and the geometry of the magnetic field are important to determine the escape process from SNRs.
Previous work about the CR escape conducted mainly two types of simulations.
One type of previous work considered a spherical shock surface and used a diffusion approximation without considering the magnetic field geometry \cite{ohira10, ptuskin03}.
The other type of previous work solved the gyration under the upstream magnetic field structure, and assumed a plane shock and an upstream escape boundary \cite{ellison96}. 
Our recent work considered a spherical shock surface and the interstellar magnetic field \cite{kamijima21}.
In that work, we solved the gyration in the upstream region to investigate CR acceleration and escape from the perpendicular shock region of typical type Ia SNRs in the ISM.
We showed that escape limits the maximum energy, which is about $10~{\rm TeV}$ for CR protons. Furthermore, the rapid perpendicular shock acceleration occurs in 20\% of the shock surface in the free expansion phase.

In this work, we investigate CR acceleration and escape from the perpendicular shock region of core-collapse SNRs by using test particle simulations.
In core-collapse supernovae, progenitors lose their mass by their stellar wind before the explosion. 
As progenitors, we consider two candidates: red supergiants (RSGs) and Wolf-Rayet (WR) stars.
For RSGs, the supernova ejecta mass and the mass blown by the RSG wind during the lifetime of RSGs are of the order of $10M_\odot$ and $1M_\odot$, respectively. 
Thus, about 10\% of the explosion energy is dissipated in the RSG wind region.
For WR stars, the mass of the free wind swept up by the SNR ejecta is much smaller than the SNR ejecta mass.
Thus, a small fraction of the explosion energy is dissipated in the free wind region of WR stars.
By considering event rates of supernova of RSGs and WR stars, these luminosities dissipated in the RSG and WR wind regions are about $10^{41}~{\rm erg\ s^{-1}}$ and $10^{39}~{\rm erg\ s^{-1}}$, respectively.
If 10\% of the energy dissipated in the wind region is converted to the CR energy, then it is sufficient to supply the Galactic CRs with energies over 1 TeV for RSGs and 1 PeV for WR stars (see Secs.~\ref{sec:stellar} and \ref{sec:condition} for details).
Therefore, we consider Galactic CRs above TeV in this work.
The magnetic field structure in the free wind region is expected to be the Parker-spiral structure with a current sheet.
The shape of the current sheet in the wind region depends on the angle between the rotation axis and the magnetic axis of progenitors.
Particles move along the current sheet due to the meandering motion because the sign of the magnetic field is inverted across the current sheet.
Therefore, the shape of the current sheet influences the escape process.
In this work, considering the shape of the current sheet, we investigate the escape-limited maximum energy for the spherical shock in the Parker-spiral magnetic field.

This paper is organized as follows. 
Simulation setups are shown in Sec.~\ref{sec:setup}.
In Secs.~\ref{sec:aligned} and \ref{sec:oblique}, we show simulation results for the case where the rotation axis and the magnetic axis are aligned (aligned rotator) and misaligned (oblique rotator), respectively.
In Sec.~\ref{sec:stellar}, we discuss the maximum energy and the luminosity of CRs in RSG winds.
The condition of acceleration to $10~{\rm TeV}$ and PeV is discussed in Sec.~\ref{sec:condition}.
Sections~\ref{sec:discussion} and \ref{sec:summary} are devoted to a discussion and summary, respectively.


\section{Simulation Setup} \label{sec:setup}

\subsection{Test particle simulation}
We perform test particle simulations to understand CR acceleration and escape in the Parker-spiral magnetic field.
In this study, we focus on protons with an energy much larger than the thermal scale.
Two types of progenitors are considered: RSGs and WR stars.
RSGs are thought to be progenitors of types II-P, II-L, and IIb supernovae.
WR stars are thought to be progenitors of types Ib and Ic supernovae.
The SNR shocks are assumed to be spherical shocks.
Particles are impulsively injected on the whole shock surface at the time when the simulation starts, $t_{\rm inj}$.
The injection time is, $t_{\rm inj}=0.3~{\rm yr}, 10~{\rm yr}, 100~{\rm yr}, 1000~{\rm yr}$ in the case of RSGs and $t_{\rm inj}=0.1~{\rm yr}, 10~{\rm yr}, 1000~{\rm yr}$ in the case of WR stars.
The initial energy of the injected particles is $1~{\rm TeV}$.
We use the particle splitting method to improve statistics of high-energy particles.

In this work, magnetic field fluctuations in the wind region (shock upstream region) and the downstream region are assumed to be zero and highly turbulent, respectively.
Instead of specifying any magnetic field distributions in the downstream region, the Bohm diffusion is uniformly assumed in the downstream region, where the downstream magnetic field is assumed to be 100 times the magnetic field strength at the shock front in the wind region. 
This corresponds to the assumption that about $1\%$ of the energy flux of the upstream flow in the shock rest frame is converted to the downstream magnetic field energy flux. 
We use different methods in the wind region and downstream region to solve the particle transport. 
In the downstream region, the random walk is solved by using the Monte Carlo method. 
Particles are isotropically scattered in the local downstream rest frame. 
The scattering angle is randomly chosen between 0 and 4$\pi$ steradian for each scattering. 
The scattering probability is given so that the scattering mean path is the gyroradius in the downstream region. 
If the downstream plasma is expanding, particles lose their energy in the shock downstream region by adiabatic loss. 
In contrast to the downstream region, we numerically solve the gyromotion in the wind region. 
The electromagnetic field in the wind region is given in the next subsection.

\subsection{Magnetic field in the wind region}
Without loss of generality, the rotation axis of progenitors is set to be the polar axis of the spherical coordinate system.
The polar and azimuthal angles, $\theta$ and $\phi$, are defined as shown in Fig.~\ref{fig:aligned}, 
where the direction of the azimuthal angle is the same as the rotational direction of the progenitor.
Hereafter, the pole means $\theta = 0, \pi$ and the equator means $\theta=\pi/2$.
In this work, as a first step, we do not consider any amplifications of the upstream magnetic field to investigate effects of the large-scale magnetic field on the escape process from the perpendicular shock region. 
The magnetic field in the wind region, $\vec{B}_{\rm w} = B_{{\rm w},r} \vec{e}_r + B_{{\rm w},\phi} \vec{e}_\phi$, is given by
\begin{eqnarray}
B_{{\rm w},r} &=& B_{\rm A} \left( \frac{R_{\rm A}}{r} \right)^2 \left\{ 1 - 2 H( \theta - \theta_{\rm CS} ) \right\} \label{eq:br} \\ 
B_{{\rm w},\phi} &=& - B_{\rm A} \frac{R_{\rm A}}{r} \frac{R_{\rm A} \Omega_{*}}{V_{\rm w}} \sin \theta \left\{ 1 - 2 H( \theta - \theta_{\rm CS} ) \right\} \label{eq:bphi} ~~,
\end{eqnarray}
where $H( \theta )$ is the Heaviside step function, $\Omega_*$ is the angular frequency of the rotation of progenitors, $R_{\rm A}$ and $B_{\rm A}$ are the Alfv\'en radius and the radial component of the magnetic field at the Alfv\'en radius.
The Alfv\'en radius is the radius at the Alfv\'en point where the magnetic field line opens from the closed dipole magnetic field line. 
The above magnetic field structure is applicable in the outer region of the Alfv\'en radius.
The Alfv\'en radius is approximately given by
\begin{eqnarray}
\frac{R_{\rm A}}{R_*} \approx 1 + \left( \eta_* + \frac{1}{4} \right)^\frac{1}{2q-2} - \left(\frac{1}{4} \right)^\frac{1}{2q-2} \label{eq:ra}~~~,
\end{eqnarray}
where $B_*$ and $R_*$ are the surface magnetic field at the equator and the radius of progenitors, respectively \cite{ud-doula08}.
In the simulations shown in Secs.~\ref{sec:aligned} and \ref{sec:oblique}, $B_*$ and $R_*$ are set to be $1~{\rm G}$ and $10^3 R_\odot$ for RSGs, and $10^3~{\rm G}$ and $5 R_\odot$ for WR stars \cite{hubrig20,hamann06}.
$\eta_* = B_*^2 R_*^2/(\dot{M} V_{\rm w})$ is the magnetic confinement parameter, and $\dot{M}$ and  $V_{\rm w}$ are the mass loss rate and the wind velocity. 
$q$ is the index about the $r$ dependency for the magnetic field inside the Alfv\'en radius.
In this work, $q=3$ because we assume the dipole magnetic field inside the Alfv\'en radius.
Then, the Alfv\'en radius is approximately given by
\begin{eqnarray}
R_{\rm A}&=& \left\{ \begin{array}{ll}
R_* & ~(~\eta_* \ll 1~) \\
R_*\eta_*^{1/4}  & ~(~\eta_* \gg 1~) \\
\end{array} \right. \\
&=& \left\{ \begin{array}{ll}
R_* & ~(~\eta_* \ll 1~) \\
R_*^{\frac{3}{2}} B_*^{\frac{1}{2}}\dot{M}^{-\frac{1}{4}} V_{\rm w}^{-\frac{1}{4}}& ~(~\eta_* \gg 1~) \\
\end{array} \right. ~~.
\label{eq:ra}
\end{eqnarray}
The strength of the radial component of the magnetic field at the Alfv\'en radius is approximately given by
\begin{eqnarray}
|B_{\rm A}|= \left\{ \begin{array}{ll}
B_* & ~(~\eta_* \ll 1~) \\
B_*^{-\frac{1}{2}} R_*^{-\frac{3}{2}} \dot{M}^{\frac{3}{4}} V_{\rm w}^{\frac{3}{4}}& ~(~\eta_* \gg 1~) \\  
\end{array} \right. ~~.
\label{eq:ba}
\end{eqnarray}
The sign of $B_{\rm A}$ is positive when the angle between the rotation axis and the magnetic axis of progenitors, $\alpha_{\rm inc}$, is below 90 degrees ($\alpha_{\rm inc} \le \pi/2$). In this case, accelerated particles drift to the equator.
On the other hand, the sign of $B_{\rm A}$ is negative when $\alpha_{\rm inc} \ge \pi/2$ and accelerated particles drift to the pole. 
The drift direction is opposite when the accelerated particles have a negative charge.

We perform simulations for  $\alpha_{\rm inc} =0, \pi/6, \pi/3, \pi/2, 2\pi/3, 5\pi/6, \pi$, where the cases of $\alpha_{\rm inc} = 0, \pi$ are called as aligned rotators and 
the cases of $\alpha_{\rm inc} \neq 0, \pi$ are called as oblique rotators. 
In oblique rotators, the current sheet has a wavy structure (see Fig.~\ref{fig:oblique}).
$\theta_{\rm CS}$ is the polar angle of the current sheet position and given by the following equation:
\begin{eqnarray}
\theta_{\rm CS} = \frac{\pi}{2} - \sin^{-1} \left[ \sin\alpha_{\rm inc} \sin \left\{ \phi + \Omega_*\left( t - \frac{ r - R_* }{V_{\rm w}} \right) \right\} \right], ~~~~
\label{eq:thetacs}
\end{eqnarray}
where $t$ is the elapsed time from the supernova explosion \citep{alanko-huotari07}.

The typical length of the wavy current sheet for oblique rotators, $\lambda = V_{\rm w} P_* = 2\pi V_{\rm w}/\Omega_*$, is given by $V_{\rm w}$ and the rotation period of progenitors, $P_*$. 
We set $P_* =40~{\rm yr}$ for RSGs, which is estimated in Betelgeuse \cite{kervalla18} and $P_*=10~{\rm days}$ for WR stars \cite{chene08}.
In this work, the current sheet and boundary between the upstream and downstream magnetic field are assumed to be sharp boundaries, 
which is valid as long as the gyroradius of high-energy particles is much larger than those boundaries.
The wind velocity, $\vec{V}_{\rm w}$, has only the radial component, $\vec{V}_{\rm w} = V_{\rm w} \vec{e}_r$. 
Then, the electric field in the wind region is given by $\vec{E}_{\rm w} = -(\vec{V}_{\rm w}/c) \times \vec{B}_{\rm w}$ in the simulation frame.
The SNR shock is a quasiparallel shock in the early phase, but as the shock propagates, the shock is a perpendicular shock [see Eqs.~(\ref{eq:br}) and (\ref{eq:bphi})].
We focus on acceleration and escape from the perpendicular shock region in this work.

\subsection{Dynamics of supernova remnants}
We consider a nonrelativistic spherical forward shock in this study. 
The time evolution of the shock velocity is given by following analytical formulas \cite{chevalier82}:
\begin{eqnarray}
u_{\rm sh} (t) &=& 
	\left\{
	\begin{array}{ll}
	\displaystyle
	\frac{n - 3}{n - 2} \left[ \frac{2}{n(n - 4)(n - 3)} \right. & \\
	\displaystyle
	\left. \times \frac{[10(n - 5)E_{\rm SN}]^{ \frac{n - 3}{2} }}{[3(n - 3)M_{\rm ej}]^{ \frac{n - 5}{2} }} \frac{V_{\rm w}}{\dot{M}t} \right]^{\frac{1}{n - 2}} & (t \le t_{\rm t}) \\
	\displaystyle
	 \sqrt{\frac{ 2 E_{\rm SN}}{M_{\rm ej}} } \left( 1 + 2\sqrt{ \frac{2 E_{\rm SN}}{M_{\rm ej}^3} } \frac{\dot{M}}{V_{\rm w}} t \right)^{-\frac{1}{2}}  & (t \ge t_{\rm t}) ~~~~,
	\end{array} 
	\right. \label{eq:ush}
\end{eqnarray}
where $E_{\rm SN} = 10^{51}~{\rm erg}, M_{\rm ej} = 5 M_\odot$ and $\dot{M} = 10^{-5} M_\odot/{\rm yr}$ are the explosion energy of supernovae, the ejecta mass, and the mass loss rate, respectively. The wind velocity is $V_{\rm w}=10^6~{\rm cm/s}$ and $10^8~{\rm cm/s}$ for RSGs and WR stars, respectively \cite{mauron11,hamann06,niedzielski02}.
In Eq.~(\ref{eq:ush}), the density profile in the wind and ejecta regions are assumed to be 
\begin{eqnarray}
\rho_{\rm w} &=& \frac{\dot{M}}{4 \pi V_{\rm w} r^2} \label{eq:rho_w} \\
\rho_{\rm ej} &\propto& 
	\left\{
	\begin{array}{ll}
	\displaystyle
	r^0 t^{-3}& ({\rm inner~ejecta}) \\
	\displaystyle
	r^{-n} t^{n-3}& ({\rm outer~ejecta})~~,
	\end{array}
	\right.
\end{eqnarray}
where $n=12$ and $10$ for RSGs and WR stars, respectively.
$r$ and $t_{\rm t}$ are the distance from the explosion center and the time when the reverse shock reaches the inner ejecta, which is given by
\begin{eqnarray}
t_{\rm t} = \frac{2}{n(n-4)(n-3)} \frac{ [3(n-3)M_{\rm ej}]^{\frac{3}{2}}}{ [10(n-5)E_{\rm ej}]^{\frac{1}{2}} } \frac{V_{\rm w}}{\dot{M}}~~~~.
\end{eqnarray}
The free wind region expands several parsecs from the explosion center in RSGs and WR stars \cite{dwarkadas05,dwarkadas07}, 
so that the SNR shock reaches the outer edge of the free wind at $t \sim 1000~{\rm yr}$. 
As the downstream velocity profile of the SNR measured in the explosion center rest frame (simulation frame), we use the approximate formula, 
\begin{equation}
u_{\rm d} (r,t) = \left( \frac{3 u_{\rm sh}(t)}{4} + \frac{V_{\rm w}}{4}\right) \left( \frac{r}{R_{\rm sh}(t)}\right),
\end{equation}
where $R_{\rm sh} = \int^t u_{\rm sh }(t') dt'$ is the shock radius.
Since the downstream is expanding (${\rm div} \vec{u_{\rm d}}>0$), particles lose their energy in the shock downstream region by adiabatic loss.

\section{Simulation Results for Aligned Rotators} \label{sec:aligned}
\begin{figure}[h]
\centering	
\includegraphics[scale=0.45]{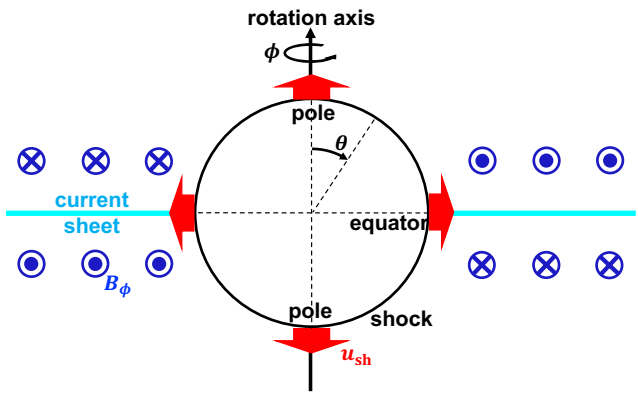}
\caption{Schematic picture of an aligned rotator ($\alpha_{\rm inc}=0$). The black circle, cyan line, and black arrow are the shock front, current sheet, and rotation axis of progenitors, respectively. The direction of the azimuthal angle, $\phi$, is the same as the rotational direction of the progenitor. The polar angle, $\theta$, is the angle measured from the rotational axis. 
\label{fig:aligned}}
\end{figure}
In this section, we perform simulations for aligned rotators ($\alpha_{\rm inc}=0, \pi$).
The schematic picture of an aligned rotator ($\alpha_{\rm inc}=0$) is shown in Fig.~\ref{fig:aligned}.
The black circle, cyan line, and black arrow are the shock front, current sheet, and rotation axis of progenitors, respectively.
For $\alpha_{\rm inc} = \pi$, $B_{{\rm w}, \phi}$ is opposite in sign to one for $\alpha_{\rm inc} = 0$.

\subsection{$\alpha_{\rm inc} = 0$ (drifting to the equator)} \label{subsec:al_eq}
\begin{figure}[h]
\centering	
\includegraphics[scale=0.86]{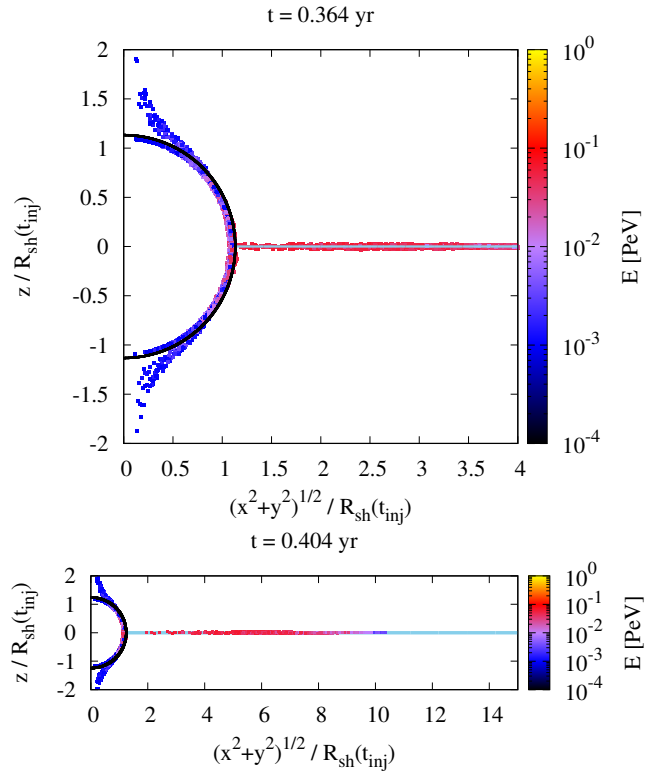}
\caption{Distribution of particles injected on the SNR surface at $t_{\rm inj} = 0.3~{\rm yr}$ for the RSG with $\alpha_{\rm inc}=0$. 
The vertical and horizontal axes are the polar-axis component of particle position and distance from the polar axis, where both the axes are normalized by the shock radius at $t_{\rm inj}$. The black hemisphere and cyan line are the shock surface and current sheet ($z=0$), respectively. The color of points means particle energy. The top and bottom panels show the distribution at $t = 0.364~{\rm yr}$ and $t = 0.404~{\rm yr}$, respectively. Note that the bottom panel shows larger scale. 
\label{fig:posi_rsg_eq_al_early}}
\end{figure}
\begin{figure}[h]
\centering	
\includegraphics[scale=0.86]{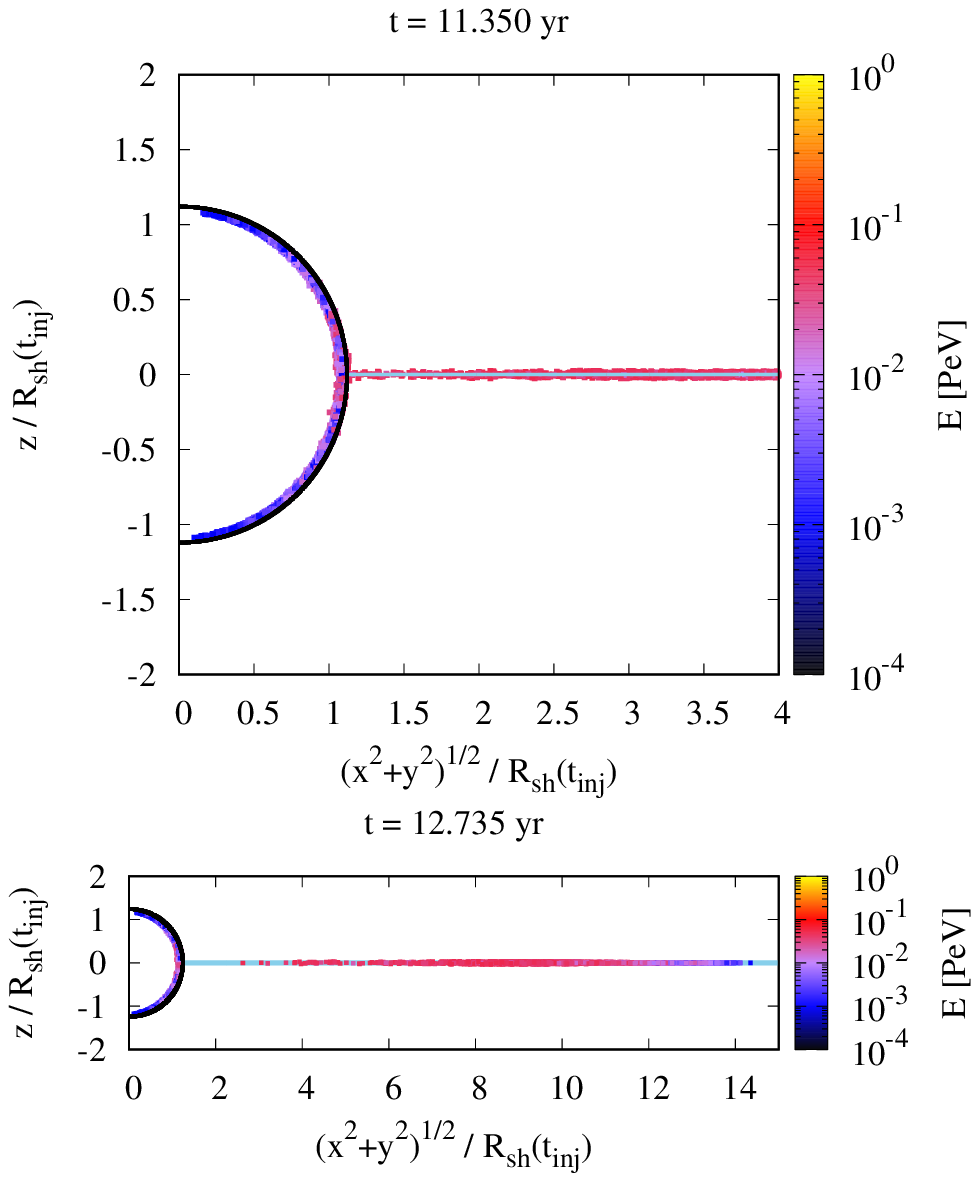}
\caption{Same as Fig.~\ref{fig:posi_rsg_eq_al_early}, but for particles injected at $t_{\rm inj} = 10~{\rm yr}$.
The top and bottom panels show the distribution at $t = 11.350~{\rm yr}$ and $t = 12.735~{\rm yr}$, respectively. Note that the bottom panel shows a larger scale. 
\label{fig:posi_rsg_eq_al}}
\end{figure}
\begin{figure}[h]
\centering	
\includegraphics[scale=0.7]{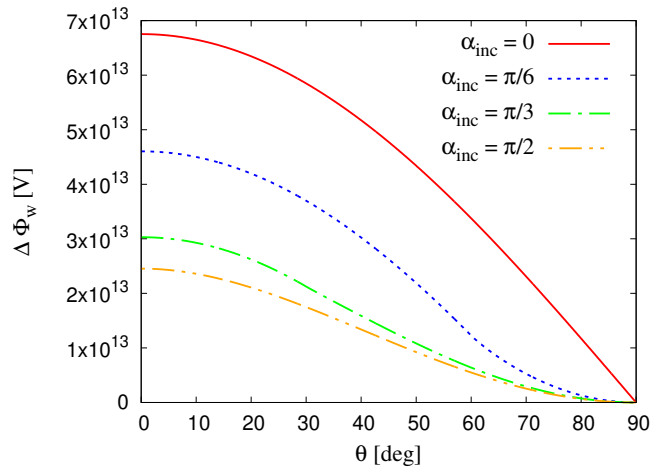}
\caption{Potential difference between the injection point with $\theta$ and the equator for RSGs with $\alpha_{\rm inc} \leq \pi/2$ at $t = 10~{\rm yr}$. The red solid, blue dotted, green dot-dashed, and orange dot-dot-dashed lines are for $\alpha_{\rm inc} = 0, \pi/6, \pi/3$, and $\pi/2$, respectively.
$\left< B_{{\rm w}, \phi} \right>$ is used to estimate the potential difference inside the wavy current sheet structure\ ($\pi/2 - \alpha_{\rm inc} \le \theta \le \pi/2$).
\label{fig:pot_rsg_eq}}
\end{figure}
\begin{figure}[h]
\centering	
\includegraphics[scale=0.62]{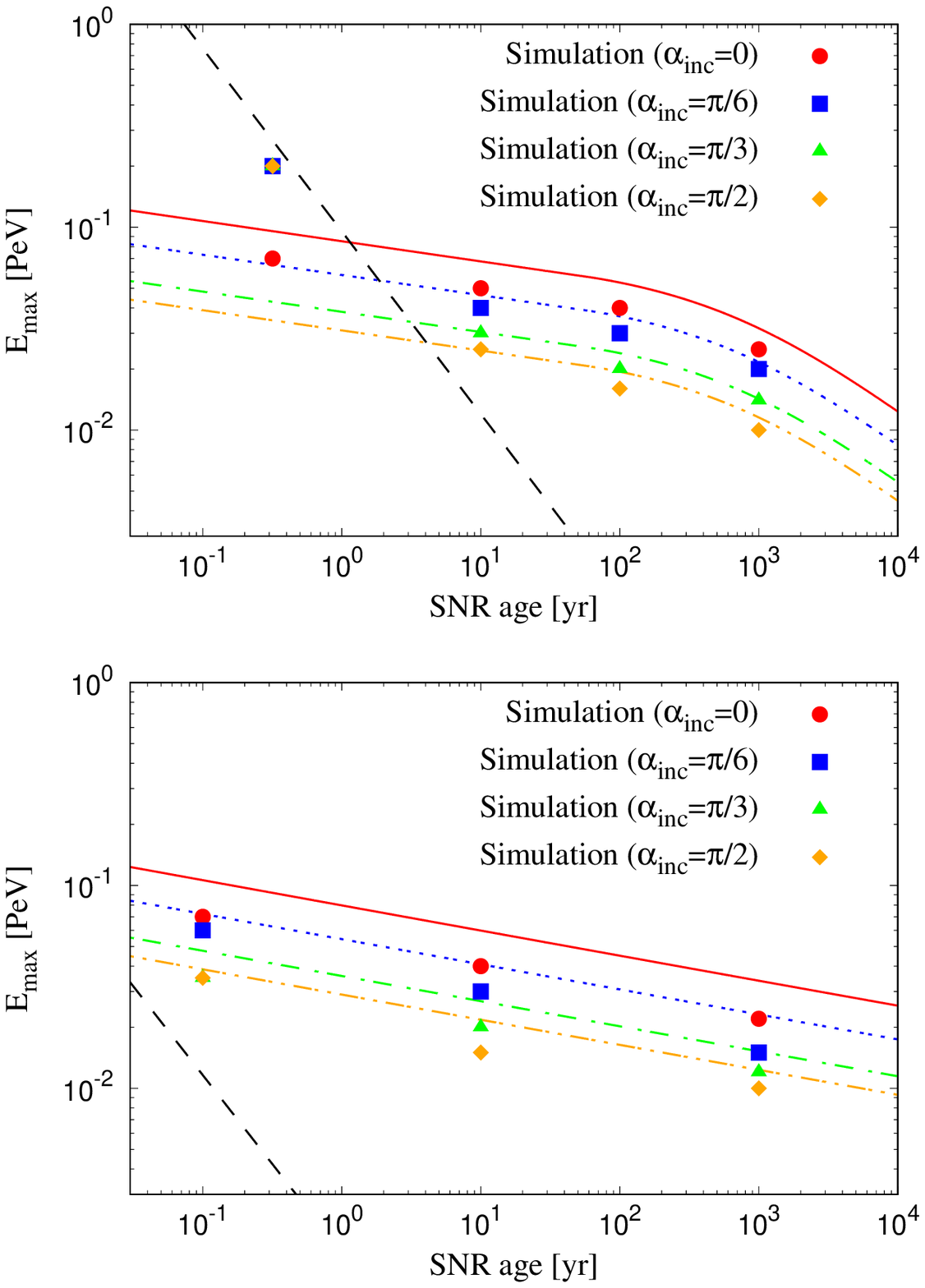}
\caption{Maximum energy as a function of the SNR age for RSGs (top) and WR stars (bottom) with $ \alpha_{\rm inc} \leq \pi/2$. 
Red filled circles, blue filled squares, green filled triangles, and orange filled diamonds are simulation results for $\alpha_{\rm inc}=0, \pi/6, \pi/3$, and $\pi/2$, respectively, where the maximum energy in simulations is estimated by the cutoff energy of $p^2 dN/dp$. The red solid, blue dotted, green dot-dashed and orange dot-dot-dashed lines are the maximum energy limited by the potential difference for $\alpha_{\rm inc}=0, \pi/6, \pi/3$, and $\pi/2$, respectively [Eq.~(\ref{eq:emax_pd_ob})]. The black dashed line is the maximum energy limited by the half wavelength of the wavy current sheet structure for oblique rotators [Eq.~(\ref{eq:half})]. 
\label{fig:emax_eq}}
\end{figure}
We first show simulation results for the RSG with $\alpha_{\rm inc}=0$.
Figure~\ref{fig:posi_rsg_eq_al_early} shows the distribution of particles injected on the SNR surface at $t_{\rm inj} = 0.3~{\rm yr}$.
The top and bottom panels are the particle distribution at $t = 0.364~{\rm yr}$ and $t =0.404~{\rm yr}$, respectively.
The vertical and horizontal axes are the polar-axis component of particle position and distance from the polar axis, where both axes are normalized by the shock radius at $t_{\rm inj}$.
The black hemisphere and cyan line are the shock surface and the current sheet ($z=0$), respectively.
The color of points means particle energy.
As one can see in Fig.~\ref{fig:posi_rsg_eq_al_early}, the particles around the poles are not accelerated and spreading from the shock surface because shocks are parallel shocks around the poles. 
However, the toroidal magnetic field, $B_{{\rm w},\phi}$, dominates over the radial magnetic field, $B_{{\rm w},r}$, as the particles propagate to the far upstream region. 
Therefore, these particles will be eventually caught up with the shock and accelerated by the shock. 
In the region of $|z|/R_{\rm sh}(t_{\rm inj}) \lesssim 0.5$, the shock is a superluminal shock, so that particles are accelerated. 
In the region of $0.5 \lesssim |z|/R_{\rm sh}(t_{\rm inj}) \lesssim 0.7$, although the shock is a subluminal shock, particles are accelerated because the particles velocity along the magnetic field line is slower than the speed of light due to a finite pitch angle.

Figure~\ref{fig:posi_rsg_eq_al} shows the distribution of particles injected on the SNR surface at $t_{\rm inj} = 10~{\rm yr}$, where the top and bottom panels show the particle distribution at $t= 11.350\ {\rm yr}$ and $t= 12.735\ {\rm yr}$, respectively. 
In contrast to the case of $t_{\rm inj} = 0.3~{\rm yr}$, most of the shock surfaces are superluminal at $t_{\rm inj} = 10~{\rm yr}$.
Therefore, except for the equator\ (the current sheet), injected particles cannot escape to the upstream region.
As one can see in Figs.~\ref{fig:posi_rsg_eq_al_early} and \ref{fig:posi_rsg_eq_al}, particles accelerated to about $0.1~{\rm PeV}$ distribute around the current sheet. 
For $\alpha_{\rm inc} = 0$, the accelerated particles with a positive charge drift on the shock surface to the equator during the DSA at the perpendicular shock. 
Once interacting with the current sheet, they start to escape from the SNR shock along the current sheet because of the meandering motion around the current sheet \cite{levy75}.

In the shock rest frame, the electric field in the wind region is 
\begin{eqnarray}
\vec{E}_{\rm w}^{\rm sh} = \frac{u_{\rm sh} - V_{\rm w}}{c} \vec{e}_r \times \vec{B}_{\rm w} \approx - \frac{u_{\rm sh}}{c} B_{{\rm w}, \phi} \vec{e}_\theta~~, 
\end{eqnarray}
where we assume the shock velocity is nonrelativistic and ignore the wind velocity because the shock velocity is much slower than the speed of light but faster than the wind velocity. 
Then, the maximum energy limited by the potential difference between the injection point with $\theta$ and the equator in the wind region is given by
\begin{eqnarray}
\varepsilon_{\rm PD} &=& e\Delta \Phi_{\rm w}  \nonumber \\
&=& e\int_\theta^{\pi/2} \left( - \frac{u_{\rm sh}}{c} B_{{\rm w}, \phi} \right) r {\rm d}\theta \nonumber \\
&=& \frac{u_{\rm sh}}{c} \frac{R_{\rm A} \Omega_*}{V_{\rm w}} eB_{\rm A} R_{\rm A} \cos{\theta} \label{eq:emax_al_eq}~~, 
\end{eqnarray}
where we considered protons as accelerated particles. 
Time dependency of the maximum energy is given through the shock velocity, $\varepsilon_{\rm PD} \propto u_{\rm sh}(t)$.
The potential difference at $t = 10~{\rm yr}$ for the RSGs is shown in Fig.~\ref{fig:pot_rsg_eq}, where the red solid line is for $\alpha_{\rm inc} = 0$.
Particles injected around the pole can be accelerated to the maximum energy in this system because of the longest drift distance from the pole to the equator.
However, it should be noted that the potential difference at $\theta = \pi/3$, which is relatively close to the equator, is one-half of the maximum potential difference at $\theta = 0$. 
Therefore, particles injected at relatively large area of the SNR surface can use most of the whole potential difference. 
Previous study also suggested that the maximum energy in the wind region is determined by the potential difference \cite{zirakashvili18}.

The time evolution of the maximum energy is shown in Fig.~\ref{fig:emax_eq}, where the top and bottom panels are for RSGs and WR stars. 
The red solid line and red filled circles show the maximum energy limited by the whole potential difference and the simulation results for the aligned rotator ($\alpha_{\rm inc}=0$). 
The maximum energy is estimated by the cutoff energy of the spectrum, $p^2 dN/dp$, in simulations. 
As one can see, the simulation results are almost in good agreement with the theoretical estimation within a factor of 2. 
The simulation results are systematically smaller than the theoretical estimation because the simulation results are average values of particles injected at various $\theta$.

In our simulations, the energy spectra of all particles injected at the later phase are in good agreement with that of the standard DSA, $dN/dp \propto p^{-2}$.
This is because simulation particles are isotropically scattered in the downstream region \cite{kamijima20}. 
Hence, the energy spectra in our simulations do not depend on the angle between the rotation axis and the magnetic axis of progenitors, $\alpha_{\rm inc}$.
On the other hand, the energy spectra are steeper than $dN/dp \propto p^{-2}$ in the earlier phase because particles escape from the polar region.

\subsection{$\alpha_{\rm inc} = \pi$ (drifting to the pole)}
\begin{figure}[h]
\centering	
\includegraphics[scale=0.5]{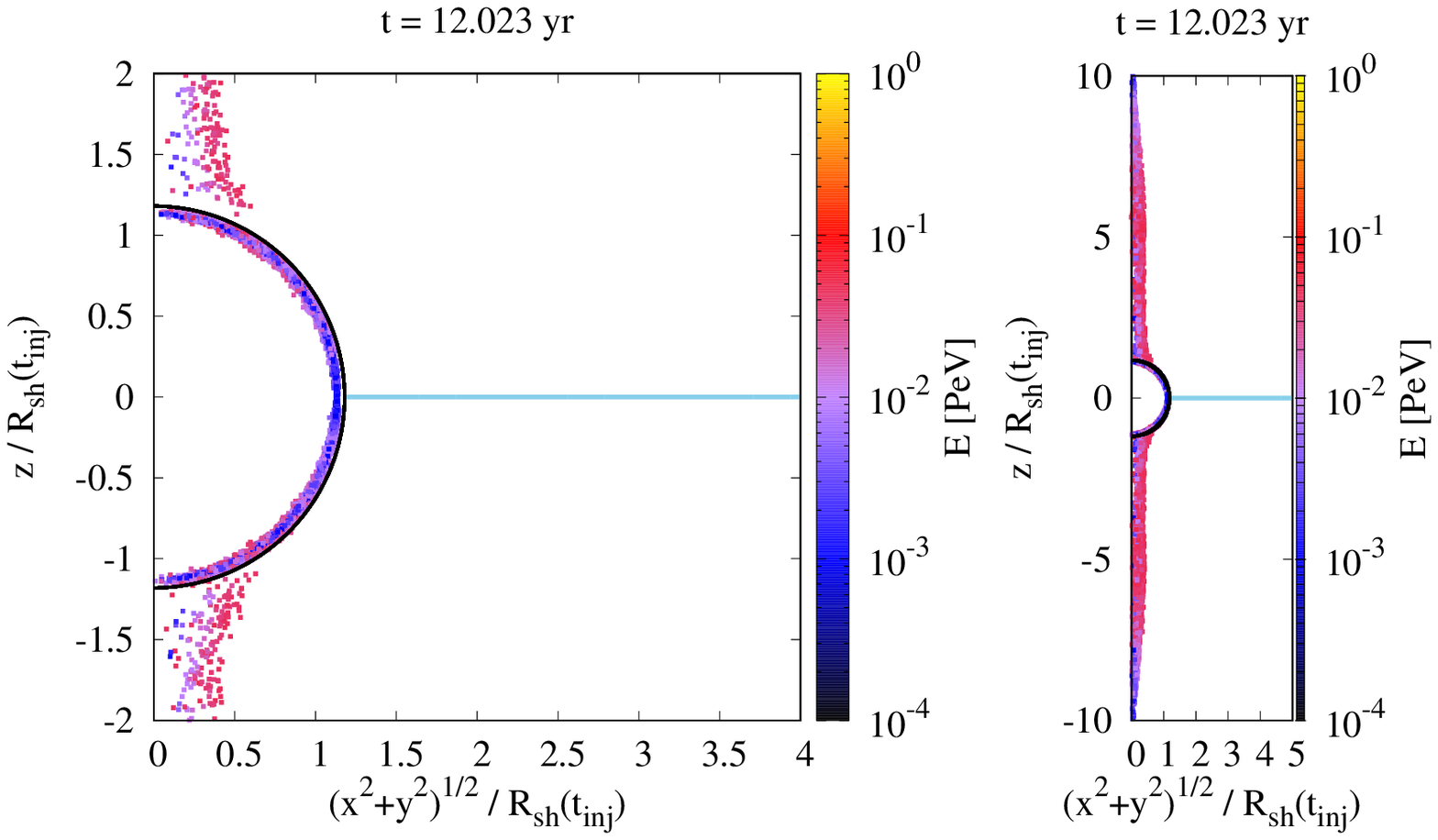}
\caption{Same as Fig.~\ref{fig:posi_rsg_eq_al}, but for $\alpha_{\rm inc}=\pi$.
Both the left and right panels show the distribution at $t = 12.023~{\rm yr}$. Note that the right panel shows larger scale. 
\label{fig:posi_rsg_pl_al}}
\end{figure}
\begin{figure}[h]
\centering	
\includegraphics[scale=0.7]{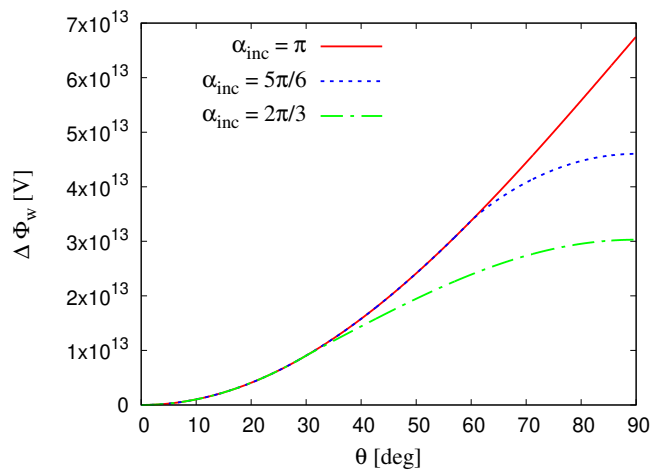}
\caption{Potential difference between the injection point with $\theta$ and the pole for RSGs with $\alpha_{\rm inc} \geq \pi/2$ at $t = 10~{\rm yr}$. The horizontal and vertical axes are $\theta$ and the potential difference, respectively.
The red solid, blue dotted and green dot-dashed lines show theoretical estimates for $\alpha_{\rm inc} = \pi, 5\pi/6$, and $2\pi/3$, respectively. 
$\left< B_{{\rm w}, \phi} \right>$ is used to estimate the potential difference inside the wavy current sheet structure\ ($\alpha_{\rm inc} - \pi/2 \le \theta \le \pi/2$).
\label{fig:pot_rsg_pl}}
\end{figure}
\begin{figure}[h]
\centering
\includegraphics[scale=0.62]{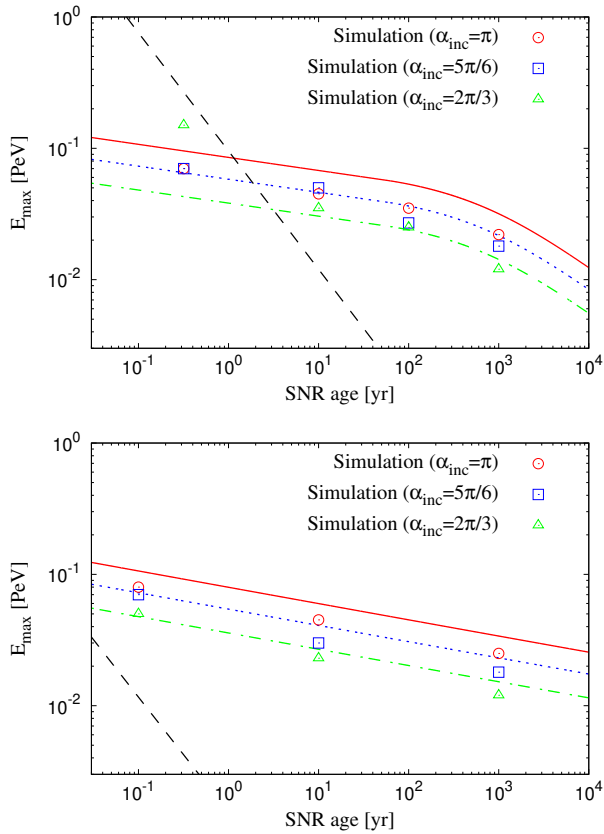}
\caption{Maximum energy as a function of the SNR age for RSGs (top) and WR stars (bottom) with $\alpha_{\rm inc} \geq \pi/2$. 
Red open circles, blue open squares, and green open triangles are simulation results for $\alpha_{\rm inc}=\pi, 5\pi/6$, and $2\pi/3$, respectively. The red solid, blue dotted and green dot-dashed lines are the maximum energy limited by the potential difference for $\alpha_{\rm inc}=\pi, 5\pi/6$, and $2\pi/3$, respectively [Eq.~(\ref{eq:emax_pot_ob_pl})]. The black dashed line is the maximum energy limited by the half wavelength of the wavy current sheet structure for oblique rotators [Eq.~(\ref{eq:half})]. 
\label{fig:emax_pl}}
\end{figure}

Next, we show simulation results for RSGs with $\alpha_{\rm inc} = \pi$.
Figure~\ref{fig:posi_rsg_pl_al} shows the distribution of particles injected on the SNR surface at $t_{\rm inj} = 10~{\rm yr}$, where the top and bottom panels show the particle distribution at $t= 12.023~{\rm yr}$.
The horizontal axis, vertical axis, and color of points are the same as in Fig.~\ref{fig:posi_rsg_eq_al_early}.
In contrast to the result for $\alpha_{\rm inc} = 0$, the accelerated particles distribute around the poles.
For $\alpha_{\rm inc} = \pi$, the accelerated particles with a positive charge drift on the shock surface to the poles. 
They escape from the poles along magnetic field lines because shocks is parallel shocks  around the poles [see Eqs.~(\ref{eq:br}) and (\ref{eq:bphi})].

As we estimated in Sec.~\ref{subsec:al_eq}, the maximum energy limited by the potential difference between the injection point with $\theta$ and the pole in the wind region is given by
\begin{eqnarray}
\varepsilon_{\rm PD} &=& e \Delta \Phi_{\rm w} \nonumber \\
&=& \int_\theta^0 \left( - \frac{u_{\rm sh}}{c} B_{{\rm w}, \phi} \right) r d\theta \nonumber \\
&=&  \frac{u_{\rm sh}}{c} \frac{R_{\rm A} \Omega_*}{V_{\rm w}} e B_{\rm A} R_{\rm A} \left( \cos{\theta} - 1 \right)~~,\label{eq:emax_al_pl}
\end{eqnarray}
where we considered protons as accelerated particles.
The potential deference for RSGs at $t = 10~{\rm yr}$ is shown in Fig.~\ref{fig:pot_rsg_pl}, where the red solid line is for $\alpha_{\rm inc} = \pi$. 
Particles injected around the equator can be accelerated to the maximum energy because of the longest drift distance from the equator to the pole. 
The maximum potential difference for $\alpha_{\rm inc} = \pi$ is the same as that for $\alpha_{\rm inc} = 0$. 
Since the potential difference at $\theta = \pi/3$ is one-half of the maximum potential difference at $\theta = \pi/2$ similarly to the case of $\alpha_{\rm inc} = 0$, particles injected at relatively large area of SNR surface can use most of the whole potential difference.

The time evolution of the maximum energy is shown in Fig.~\ref{fig:emax_pl}, where the top and bottom panels are for RSGs and WR stars, respectively.
The red solid line and red open circles show the maximum energy limited by the whole potential difference and simulation results for the aligned rotator with $\alpha_{\rm inc}=\pi$, respectively.
The simulation results for $\alpha_{\rm inc}=\pi$ are almost the same as that for $\alpha_{\rm inc}=0$ and 
almost in good agreement with the theoretical estimation.

\section{Simulation results for oblique rotators} \label{sec:oblique}

\begin{figure}[h]
\centering	
\includegraphics[scale=0.45]{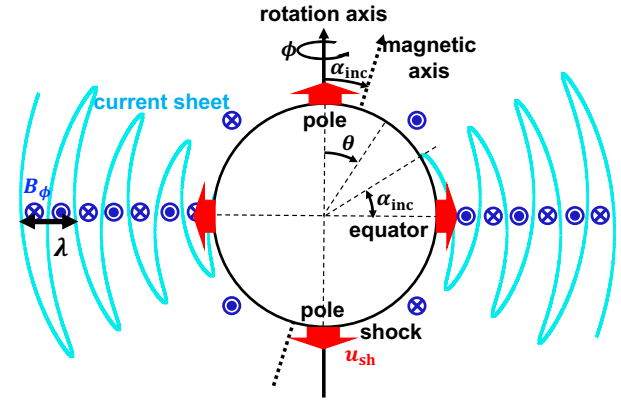}
\caption{Schematic picture of an oblique rotator~($\alpha_{\rm inc} \le \pi/2$). The black circle, cyan line, black solid arrow, and black dotted arrow are the shock front, current sheet, rotation axis, and magnetic axis of progenitors, respectively. The direction of the azimuthal angle, $\phi$, is the same as the rotational direction of the progenitor. The polar angle, $\theta$, is the angle measured from the rotational axis. $\lambda$ is the typical length scale of the wavy current sheet.
\label{fig:oblique}}
\end{figure}
\begin{figure}[h]
\centering	
\includegraphics[scale=0.7]{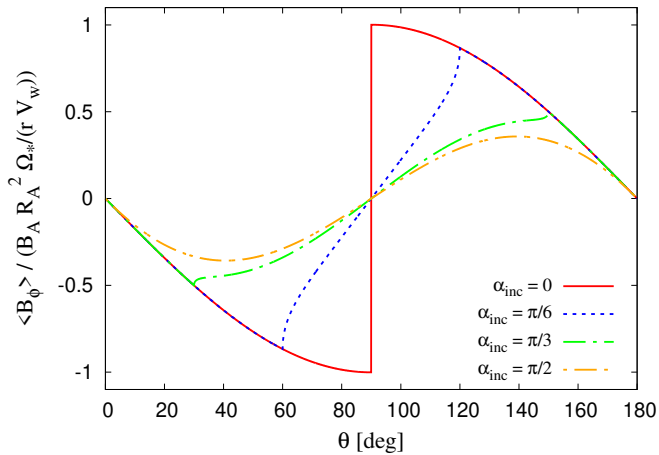}
\caption{$\phi$-averaged $B_{{\rm w},\phi}$, $\left< B_{{\rm w},\phi} \right>$. 
The horizontal and vertical axes show $\theta$ and $\left< B_{{\rm w}, \phi} \right>$ normalized by $B_{\rm A}R_{\rm A}^2\Omega_{*}/(rV_{*})$. The red solid line, blue dotted line, green dot-dashed line and orange dot-dot-dashed line are for $\alpha_{\rm inc} = 0, \pi/6, \pi/3$, and $\pi/2$, respectively.
\label{fig:bphi}}
\end{figure}
In this section, we perform simulations for oblique rotators ($\alpha_{\rm inc} \neq 0, \pi$). 
The schematic picture of an oblique rotator ($\alpha_{\rm inc} \le \pi/2$) is shown in Fig.~\ref{fig:oblique}.
The black circle and cyan wavy line are the shock front and current sheet, respectively.
The black solid and dashed arrows are the rotation and magnetic axes of progenitors, respectively.
For $\alpha_{\rm inc} \ge \pi/2$, $B_{{\rm w},\phi}$ is opposite in sign to one for $\alpha_{\rm inc} \le \pi/2$.
The magnetic field structure outside the wavy current sheet region is the same as that of aligned rotators.
On the other hand, in the wavy current sheet region, the magnetic field reverses its sign every one wavelength of the wavy current sheet, $\lambda$, along the radial direction. 
Even though we take the spatial average of the magnetic field over $\lambda$, it has a finite value except for $\theta=\pi/2$.
Therefore, particles with a larger gyroradius than $\lambda$ feel the mean magnetic field, $\left< \vec{B} \right>$.
To estimate the mean magnetic field, we have to take the spatial average over the gyroradius, but it can be replaced by the average over the azimuthal angle, $\phi$, because the purpose of taking the spatial average is to estimate the asymmetry of the reversal structure of the magnetic field. 
Here, we consider the case for $\alpha_{\rm inc} \le \pi/2$ and the azimuthal component of magnetic field, $B_{{\rm w},\phi}$ because we focus on regions where the radial component is negligible. 
$\left< B_{{\rm w},\phi} \right>$ for $\alpha_{\rm inc} \ge \pi/2$ is the opposite in sign to one for  $\alpha_{\rm inc} \le \pi/2$.
At a point in the wavy current sheet region $(r, \theta)$, the current sheet is located at $\phi=\phi_{+}$ and $\phi_{-}$.  From Eq.~(\ref{eq:thetacs}), $\phi_{\pm}$ are given by
\begin{eqnarray}
\phi_{\pm} = \frac{\pi}{2} - \Omega_*\left( t - \frac{r - R_*}{V_{\rm w}} \right) \pm \cos^{-1} \left( \frac{\cos \theta}{\sin \alpha_{\rm inc}} \right)~~. 
\end{eqnarray}
Then, $\left< B_{{\rm w},\phi} \right>$ is calculated as follows:
\begin{eqnarray}
\left< B_{{\rm w},\phi} \right> &\approx& \frac{1}{2\pi} \int_{\phi_-}^{2\pi + \phi_-} d\phi B_{{\rm w},\phi} \nonumber \\
&=& - \frac{B_{\rm A}}{2\pi} \frac{R_{\rm A}}{r}   \frac{R_{\rm A} \Omega_{*}}{V_{\rm w}}  \sin \theta  \left\{ - \int_{\phi_-}^{\phi_+} d\phi + \int_{\phi_+}^{2\pi + \phi_-} d\phi \right\} \nonumber \\
&=& - B_{\rm A} \frac{R_{\rm A}}{r} \frac{R_{\rm A} \Omega_{*}}{V_{\rm w}}  \sin \theta \left\{ 1- \frac{2}{\pi} \cos^{-1} \left( \frac{\cos \theta}{\sin \alpha_{\rm inc}} \right) \right\} 
\label{eq:mean_bphi} ~~~~~~.
\end{eqnarray}
This is valid in the wavy current sheet region ($\pi/2 - \alpha_{\rm inc} \le \theta \le \pi/2 +\alpha_{\rm inc}$), but $\left< B_{{\rm w},\phi} \right>=B_{{\rm w},\phi}$ in other regions. 
Figure~\ref{fig:bphi} shows $\left< B_{{\rm w}, \phi} \right>$ as a function of $\theta$ for $\alpha_{\rm inc} = 0, \pi/6, \pi/3$, and $\pi/2$. 
$\left< B_{{\rm w}, \phi} \right>=B_{{\rm w}, \phi}$ for aligned rotators (red solid line), but $\left< B_{{\rm w}, \phi} \right> < B_{{\rm w}, \phi}$ in the wavy current region for oblique rotators (the other lines). 
As we already mentioned, $\left< B_{{\rm w},\phi} \right>$ is not zero except for $\theta=\pi/2$, so that particles with a gyroradius larger than $\lambda$ drift on the shock surface due to the mean magnetic field. 
However, $\left< B_{{\rm w},\phi} \right>=0$ at $\theta = \pi/2$ and its sign changes across $\theta = \pi/2$ even for oblique rotators, so that particles with a gyroradius larger than $\lambda$ can escape to the far upstream region along the equatorial plane. 
It should be noted that particles with any energies can escape along the equatorial plane for the aligned rotator,  but only sufficiently high-energy particles can escape for the oblique rotator.

\subsection{$\alpha_{\rm inc} \le \pi/2$ (drifting to the equator)} \label{sec:ob_eq}
\begin{figure}[h]
\centering	
\includegraphics[scale=0.86]{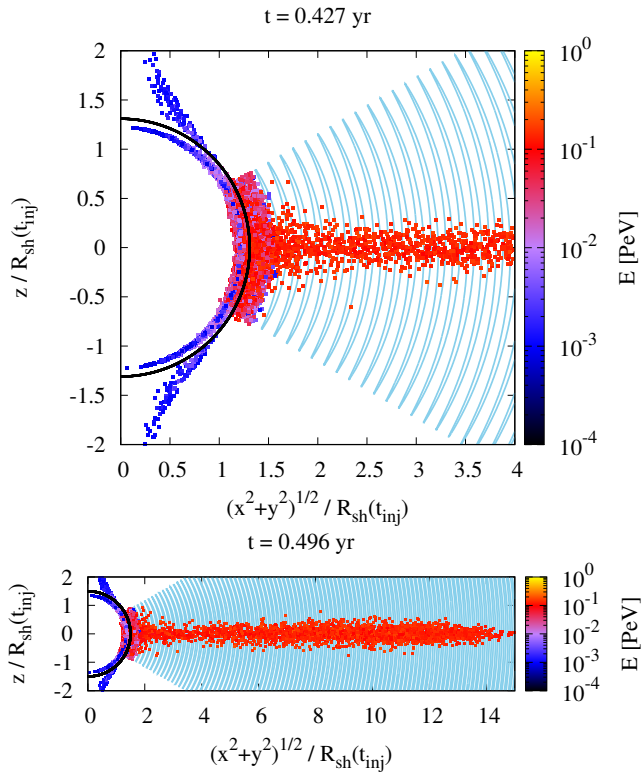}
\caption{Distribution of particles injected on the SNR surface at $t_{\rm inj} = 0.3~{\rm yr}$ for the RSG with $\alpha_{\rm inc}=\pi/6$. The vertical and horizontal axes are the same as in Fig.~\ref{fig:posi_rsg_eq_al_early}. The black hemisphere and cyan line are the shock surface and wavy current sheet in the plane~($y = 0$ and $x>0$), respectively. The color of points means particle energy. The top and bottom panels show the distribution at $t = 0.427~{\rm yr}$ and $t = 0.496~{\rm yr}$, respectively. Note that the bottom panel shows larger scale. 
\label{fig:posi_rsg_eq_ob_e}}
\end{figure}
\begin{figure}[h]
\centering	
\includegraphics[scale=0.8]{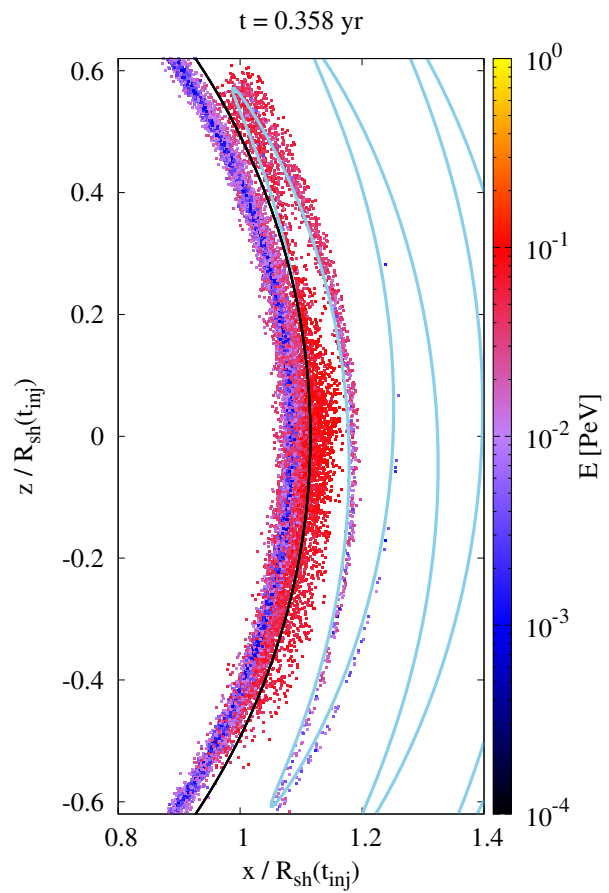}
\caption{Distribution of particles injected on the SNR surface at $t_{\rm inj} = 0.3~{\rm yr}$ for the RSG with $\alpha_{\rm inc}=\pi/6$. The vertical and horizontal axes are the $z$ and $x$ of the particle position, where both axes are normalized by the shock radius at $t_{\rm inj}$. The black hemisphere and cyan line are the shock surface and wavy current sheet in the plane~($y = 0$ and $x>0$), respectively. The color of points means particle energy. 
This shows the distribution of particles located in $-0.05 \le y/R_{\rm sh} \le 0.05$ at $t = 0.358~{\rm yr}$. 
\label{fig:meandering}}
\end{figure}
\begin{figure}[h]
\centering	
\includegraphics[scale=0.86]{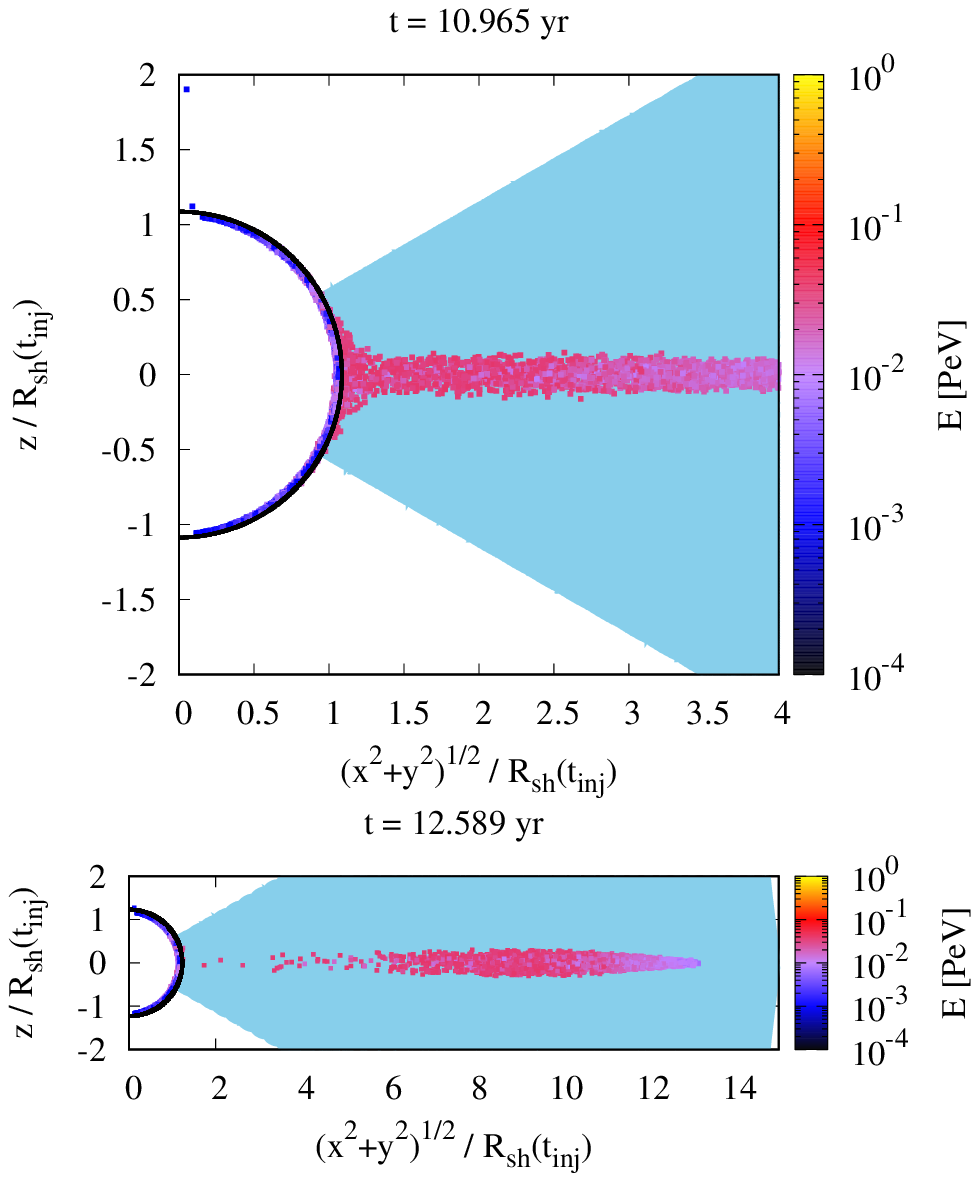}
\caption{Same as Fig.~\ref{fig:posi_rsg_eq_ob_e}, but for particles injected at $t_{\rm inj} = 10~{\rm yr}$.
The top and bottom panels show the distribution at $t = 10.965~{\rm yr}$ and $t = 12.589~{\rm yr}$, respectively. Note that the bottom panel shows larger scale. 
\label{fig:posi_rsg_eq_ob_l}}
\end{figure}
In this subsection, we show the simulation result for $\alpha_{\rm inc}=\pi/6$ as a typical example of $\alpha_{\rm inc} \le \pi/2$.
Figure~\ref{fig:posi_rsg_eq_ob_e} shows the distribution of particles injected on the SNR surface at $t_{\rm inj} = 0.3~{\rm yr}$, where the top and bottom panels show the particle distribution at $t= 0.427\ {\rm yr}$ and $t= 0.496\ {\rm yr}$, respectively. 
The horizontal axis, vertical axis and color of points are the same as in Fig.~\ref{fig:posi_rsg_eq_al_early}.
The black hemisphere and wavy cyan line are the shock surface and current sheet in the plane~($y=0$ and $x>0$), respectively.
In the early phase ($t\lesssim 0.5 {\rm yr}$), most regions on the shock front are subluminal because the radial magnetic field is stronger than the toroidal magnetic field. 
However, the simulation shows that some particles are accelerated in the wavy current sheet region. 
In the shock upstream region, particles are distributed mainly in three regions, the polar, wavy current sheet, and equatorial plane regions. 

In the polar region ($\theta < \pi/3$ and $\theta > 2\pi/3$), the particles are not accelerated and escape from the shock while moving along the spiral magnetic field line because the shock in this region is a subluminal shock at $t \lesssim 0.5 {\rm yr}$.
However, these particles will be eventually caught up with the shock later 
because the toroidal component of the magnetic field dominates over the radial component as the particles propagate to the far upstream region. 
This behavior is the same as the case for $\alpha_{\rm inc} = 0$ shown in Fig.~\ref{fig:posi_rsg_eq_al_early}.

In the wavy current sheet structure\ ($\pi/3 \le \theta \le 2\pi/3$), the particles are accelerated and trapped around the shock except for particles in the equatorial plane. 
At $t = 0.427\ {\rm yr}$, the shock is a superluminal shock inside the wavy current sheet structure other than the current sheet region.
Accelerating particles drift on the shock surface until interacting with the current sheet.
Once particles with a gyroradius smaller than the half wavelength of the wavy current sheet interact with the current sheet, these particles propagate along the wavy current sheet due to the meandering motion. 
To see this propagation more clearly, we show the distribution of particles at $t = 0.358~{\rm yr}$ in Fig.~\ref{fig:meandering}, where
we show only particles located in $-0.05 \le y/R_{\rm sh} \le 0.05$.
As one can see in Fig.~\ref{fig:meandering}, the particles propagate along the wavy current sheet.
The radial velocity of the particles, $c\lambda/(r \sin \alpha_{\rm inc})$, is not so fast even if particles move along the current sheet with the speed of light.
Since the ratio of $r \sin \alpha_{\rm inc}$ and $\lambda$ becomes large as particles escape to the upstream region, the radial velocity eventually becomes slower than the shock velocity, so that the particles are caught up with the shock and accelerated by the shock. 
In contrast to the case of $\alpha_{\rm inc} = 0$, owing to the wavy current sheet, particles are trapped around the shock. 
Then, the maximum energy can be larger than the potential difference between the pole and equator even though the upstream magnetic field is not amplified.

Once the gyroradius of accelerated particles becomes larger than the half wavelength of the wavy current sheet, 
the particles cannot propagate along the current sheet 
because the particles pass more than two current sheets during one gyromotion.  
Then, the particles feel the mean magnetic field structure estimated in Eq.~(\ref{eq:mean_bphi}), so that particles drift to the equator and escape to the far upstream region. 
As a result, the maximum energy in the early phase is limited by the half wavelength of the wavy current sheet.  
From the condition of $r_{\rm g} = \lambda/2$, the maximum energy is given by
\begin{equation}
\varepsilon_{{\rm max}, \lambda/2} = \frac{\pi e |B_{\rm A}| R_{\rm A}^2 }{R_{\rm sh}}, ~~\label{eq:half}
\end{equation}
where $|B_A|$ and $R_A$ are the magnetic field strength and radius at the Alfv\'en point.
This does not depend on the angle between the rotation axis and the magnetic axis of progenitors, $\alpha_{\rm inc}$.
The gyroradius becomes smaller in the earlier phase because $B_{{\rm w},\phi}$ is inversely proportional to $R_{\rm sh}$.
On the other hand, $\lambda$ is constant, so that the maximum energy becomes larger in the earlier phase.

For RSGs, the maximum energy is estimated as follows
\begin{eqnarray}
\varepsilon_{{\rm max}, \lambda/2} &\approx& 1~{\rm TeV} \left( \frac{R_{\rm sh}}{1~{\rm pc}} \right)^{-1} \left( \frac{B_*}{1~{\rm G}} \right)^{\frac{1}{2}} \left( \frac{R_*}{10^3 R_\odot} \right)^{\frac{3}{2}} \nonumber \\
&&\times \left( \frac{\dot{M}}{10^{-5}M_\odot/{\rm yr}} \right)^{\frac{1}{4}} \left( \frac{V_{\rm w}}{10^6~{\rm cm/s}} \right)^{\frac{1}{4}}
~~,
\end{eqnarray}
where $\eta_* \approx 7.4$ is used and $R_{\rm A}$ and $B_{\rm A}$ are given by Eqs.~(\ref{eq:ra}) and (\ref{eq:ba}).
For WR stars, $\varepsilon_{{\rm max}, \lambda/2}$ is about $3.5 \times 10^{-2}~{\rm TeV}$ when $R_{\rm sh}$ and $\eta_*$ are about $1~{\rm pc}$ and $1.8$, respectively.
$\varepsilon_{{\rm max}, \lambda/2}$ is larger in the earlier phase because of the smaller $R_{\rm sh}$ in the earlier phase.

Figure~\ref{fig:posi_rsg_eq_ob_l} shows the distribution of particles injected on the SNR surface at $t_{\rm inj} = 10~{\rm yr}$, where the top and bottom panels show the particle distribution at $t= 10.965\ {\rm yr}$ and $t= 12.589\ {\rm yr}$, respectively. 
In contract to the result for $t_{\rm inj} = 0.3~{\rm yr}$, in the upstream region inside the wavy current sheet structure, we can see only escaping particles moving along the equator. 
Particles cannot propagate to the upstream region along the wavy current sheet because the path along the wavy current sheet becomes longer at the later phase. 
Therefore, particles cannot escape from the shock front except for in the equatorial plane where particles with $r_{\rm g}> \lambda/2$ feel an effectively zero magnetic field, $\left< B_{{\rm w},\phi} \right>\approx 0$.
The particle energy is limited by the potential difference between the injection position with $\theta$ and the equator, $\Delta \Phi_{\rm w}$.
Here, we consider $0 \le \theta \le \pi/2$ and particles with $r_{\rm g} > \lambda/2$ to estimate the upstream electric potential in the shock rest frame, $\Phi_{\rm w}$. 
In the shock rest frame, the electric field 
is given by $\vec{E}_{\rm w}^{\rm sh} \approx -(u_{\rm sh}/c) \left< B_{{\rm w},\phi} \right> \vec{e}_\theta$ in the upstream region, where $\left< B_{{\rm w},\phi} \right> = B_{{\rm w},\phi}$ outside the wavy current sheet structure and $\left< B_{{\rm w},\phi} \right>$ is given by Eq.~(\ref{eq:mean_bphi}) inside the wavy current sheet structure. Then, the electric potential, $\Phi_{\rm w} = \int_0^{\theta} d\theta' \left( -r E_{\rm w}^{\rm sh} \right)$, 
outside the wavy current sheet structure, $0 \le \theta \le \pi/2 - \alpha_{\rm inc}$, is
\begin{eqnarray}
\Phi_{\rm w} = \Phi_{\rm w,0} + \frac{u_{\rm sh}}{c} \frac{R_{\rm A} \Omega_*}{V_{\rm w}} B_{\rm A} R_{\rm A} \cos \theta \label{eq:phi_w_out}~~,
\end{eqnarray}
where $\Phi_{\rm w,0}$ is the electric potential at the pole.
The electric potential inside the wavy current sheet structure, $\pi/2 - \alpha_{\rm inc} \le \theta \le \pi/2$, is 
\begin{eqnarray}
\Phi_{\rm w} &=& \Phi_{\rm w,0} + \frac{u_{\rm sh}}{c} \frac{R_{\rm A} \Omega_*}{V_{\rm w}} B_{\rm A} R_{\rm A} \nonumber \\
&& \times \left[ \cos \theta - \frac{2}{\pi} \cos \theta \cos^{-1} \left( \frac{\cos \theta}{\sin \alpha_{\rm inc}} \right) \right. \nonumber \\
&&\left. + \frac{2}{\pi} \sqrt{\sin^2 \alpha_{\rm inc} - \cos^2 \theta} \right] \label{eq:phi_w_in}~~.
\end{eqnarray}

If particles are injected outside the wavy current sheet structure, $0 \le \theta \le \pi/2 - \alpha_{\rm inc}$, the potential difference between an injection position with $\theta$ and the equator, $\Delta \Phi_{\rm w} = \Phi_{\rm w}(\theta) - \Phi_{\rm w}(\pi/2)$, is 
\begin{eqnarray}
\Delta \Phi_{\rm w} = \frac{u_{\rm sh}}{c} \frac{R_{\rm A} \Omega_*}{V_{\rm w}} B_{\rm A} R_{\rm A}  \left( \cos \theta - \frac{2}{\pi} \sin \alpha_{\rm inc} \right)~~. \label{eq:pot_eq_ob1}
\end{eqnarray}
On the other hand, if particles are injected inside the wavy current sheet, $\pi/2 - \alpha_{\rm inc} \le \theta \le \pi/2$, the potential difference, $\Delta \Phi_{\rm w} = \Phi_{\rm w}(\theta) - \Phi_{\rm w}(\pi/2)$, is 
\begin{eqnarray}
\Delta \Phi_{\rm w} &=& \frac{u_{\rm sh}}{c} \frac{R_{\rm A} \Omega_*}{V_{\rm w}} B_{\rm A} R_{\rm A} \nonumber \\
&&\times \left[ \cos \theta - \frac{2}{\pi} \cos \theta \cos^{-1} \left( \frac{\cos \theta}{\sin \alpha_{\rm inc}} \right) \right. \nonumber \\
&& \left. + \frac{2}{\pi} \sqrt{\sin^2 \alpha_{\rm inc} - \cos^2 \theta} - \frac{2}{\pi} \sin \alpha_{\rm inc} \right].~~\label{eq:pot_eq_ob2}
\end{eqnarray}
The potential difference for oblique rotators is shown in Fig.~\ref{fig:pot_rsg_eq}.
The blue dotted, green dot-dashed, and orange dot-dot-dashed lines are the potential difference for $\alpha_{\rm inc} = \pi/6, \pi/3$, and $\pi/2$, respectively.
The potential difference decreases as $\alpha_{\rm inc}$ approaches to $\pi/2$ because the wavy current region increases and the mean magnetic field, $\left< B_{{\rm w},\phi} \right>$, decreases.
The maximum energy limited by the potential difference, $\varepsilon_{\rm max,PD}$, is
\begin{eqnarray}
\varepsilon_{\rm max,PD} &=& e \Delta \Phi_{\rm w}(\theta = 0) \nonumber \\
&=&  \left( 1 - \frac{2}{\pi} \sin \alpha_{\rm inc} \right) \frac{u_{\rm sh}}{c} \frac{R_{\rm A} \Omega_*}{V_{\rm w}} e B_{\rm A} R_{\rm A} \label{eq:emax_pd_ob}~~.
\end{eqnarray}
The difference from aligned rotators ($\alpha_{\rm inc}=0,\pi$) is only the factor of $1 - (2/\pi) \sin \alpha_{\rm inc}\gtrsim 0.37$. 
For RSGs, the potential-limited maximum energy is estimated to be
\begin{eqnarray}
\varepsilon_{\rm max,PD} &\approx& 44~{\rm TeV} \left( 1 - \frac{2}{\pi} \sin \alpha_{\rm inc} \right) \left( \frac{u_{\rm sh}}{10^{-2}c} \right)  \nonumber \\
&&\times \left( \frac{B_*}{1~{\rm G}} \right)^{\frac{1}{2}} \left( \frac{R_*}{10^3 R_\odot} \right)^{\frac{3}{2}} \left( \frac{\dot{M}}{10^{-5}M_\odot/{\rm yr}} \right)^{\frac{1}{4}} \nonumber \\
&&\times \left( \frac{V_{\rm w}}{10^6~{\rm cm/s}} \right)^{-\frac{3}{4}} \left( \frac{P_*}{40~{\rm yr}} \right)^{-1} \label{eq:emax_rsg_pd_eta>1}~~, 
\end{eqnarray}
where $\eta_* \approx 7.4$ is used and $R_{\rm A}$ and $B_{\rm A}$ are given by Eqs.~(\ref{eq:ra}) and (\ref{eq:ba}).
If $\eta_* \ll 1$, then the Alfv\'en radius and magnetic field at the Alfv\'en radius are $R_{\rm A}=R_*$ and $B_{\rm A}=B_*$, respectively.
For WR stars, $\varepsilon_{\rm max,PD}$ is about $23~{\rm TeV}$ when the shock velocity is $u_{\rm sh}=10^{-2}c$.
The maximum energy for oblique rotators is shown in Fig.~\ref{fig:emax_eq}.
The blue dotted, green dot-dashed, and orange dot-dot-dashed lines are the maximum energy limited by the whole potential difference for $\alpha_{\rm inc} = \pi/6, \pi/3$, and $\pi/2$, respectively.
The blue filled squares, green filled triangles, and orange filled diamonds are the simulation results for $\alpha_{\rm inc} = \pi/6, \pi/3$, and $\pi/2$, respectively.
The black dashed line is the maximum energy limited by the half wavelength of the wavy current sheet, $\varepsilon_{{\rm max}, \lambda/2}$.
As one can see in Fig.~\ref{fig:emax_eq}, the simulation results are almost in good agreement with the theoretical estimation within a factor of 2.
For RSGs, the maximum energy at $t_{\rm inj} = 0.3~{\rm yr}$ is limited by the half wavelength of the wavy current sheet, $\varepsilon_{{\rm max}, \lambda/2}$, which does not depend on the angle between the rotation axis and the magnetic axis of progenitors, $\alpha_{\rm inc}$.
Since particles are trapped by the wavy current sheet around the shock surface, 
particles can be accelerated for a longer time than during the drifting time from the pole to the equator.
Therefore, $\varepsilon_{{\rm max}, \lambda/2}$ can be larger than $\varepsilon_{\rm max,PD}$ in the early phase. 
However, $\varepsilon_{{\rm max}, \lambda/2}$ quickly decreases with time. 
The magnetic field strength is inversely proportional to the shock radius but the wave length of the wavy current sheet is constant with the shock radius. 
Thus, the maximum energy given by $r_{\rm g} = \lambda/2$ becomes smaller than $\varepsilon_{\rm max,PD}$ in the later phase. Then, the 
maximum energy is limited by the potential difference in the later phase.
For WR stars, compared with RSGs, $\varepsilon_{{\rm max}, \lambda/2}$ is larger than the potential difference only in the very early phase, and the maximum energy is limited by the whole potential difference in almost all age of the SNR.

\subsection{$\alpha_{\rm inc} \ge \pi/2$ (drifting to the pole)} \label{sec:ob_pl}
\begin{figure}[h]
\centering	
\includegraphics[scale=0.5]{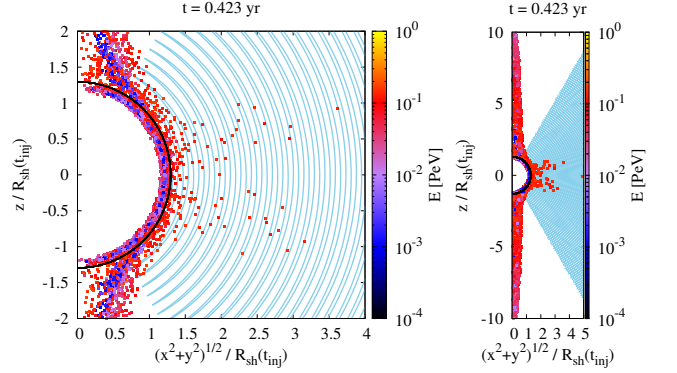}
\caption{Same as Fig.~\ref{fig:posi_rsg_eq_ob_e}, but for $\alpha_{\rm inc} = 2\pi/3$. Both the left and right panels show the distribution at $t = 0.423~{\rm yr}$, but the right panel shows larger scale.
\label{fig:posi_rsg_pl_ob_e}}
\end{figure}
\begin{figure}[h]
\centering	
\includegraphics[scale=0.5]{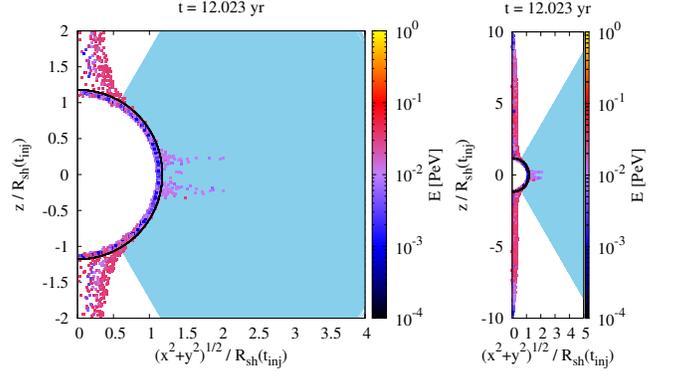}
\caption{Same as Fig.~\ref{fig:posi_rsg_pl_ob_e}, but for particles injected at $t_{\rm inj} = 10~{\rm yr}$.
Both the left and right panels show the distribution at $t = 12.023~{\rm yr}$, but the right panel shows larger scale. 
\label{fig:posi_rsg_pl_ob_l}}
\end{figure}
In this subsection, first of all, we show the simulation result for $\alpha_{\rm inc} = 2\pi/3$ as an example of $\alpha_{\rm inc} \ge \pi/2$. 
Figure~\ref{fig:posi_rsg_pl_ob_e} shows the distribution of particles injected on the SNR surface at $t_{\rm inj} = 0.3~{\rm yr}$, where both the left and right panels show the particle distribution at $t= 0.423\ {\rm yr}$. 
Injected particles are accelerated while drifting to the poles. 
After reaching the poles, the accelerated particles escape to the far upstream region while moving along the poles.
As one can see in Fig.~\ref{fig:posi_rsg_pl_ob_e}, in comparison with the case of $\alpha_{\rm inc} \le \pi/2$, 
a few particles are distributed more sparsely in the upstream region inside the wavy current sheet structure.
This is because, for $\alpha_{\rm inc} \ge \pi/2$, the accelerated particles drift to the poles every shock crossing rather than to the equator. 
The maximum energy is limited by the half wavelength of the wavy current sheet [see Eq.~(\ref{eq:half})].

Figure~\ref{fig:posi_rsg_pl_ob_l} shows the distribution of particles injected on the SNR surface at $t_{\rm inj} = 10~{\rm yr}$, where both left and right panels show the particle distribution at $t= 12.023\ {\rm yr}$. 
The distribution of particles inside the wavy current sheet is similar to that for $t_{\rm inj}=0.3~{\rm yr}$, but the particle energy is different from that for $t_{\rm inj}=0.3~{\rm yr}$. 
This is because the maximum energy given by $r_{\rm g} = \lambda/2$ becomes smaller than that for $t_{\rm inj}=0.3~{\rm yr}$. 
Accelerated particles with $r_{\rm g} \ge \lambda/2$ cannot propagate along the wavy current sheet by the meandering motion. 
These particles feel the mean magnetic field, $\left< B_{{\rm w},\phi} \right>$, while drifting to the poles. 
Then, the maximum energy is limited by the potential difference between the injection position with $\theta$ and the pole. 
As we mentioned in Sec.~\ref{sec:ob_eq}, the potential structure depends on whether particles are injected inside or outside the wavy current sheet structure. 
Here, we consider particles injected in $0 \le \theta \le \pi/2$.
If particles are injected outside the wavy current sheet structure~($0 \le \theta \le \alpha_{\rm inc} - \pi/2$), from Eq.~(\ref{eq:phi_w_out}), then the potential difference, $\Delta \Phi_{\rm w} = \Phi_{\rm w}(\theta) - \Phi_{\rm w}(0)$, is
\begin{eqnarray}
\Delta \Phi_{\rm w} = \frac{u_{\rm sh}}{c} \frac{R_{\rm A} \Omega_*}{V_{\rm w}} B_{\rm A} R_{\rm A}  \left( \cos \theta - 1\right) \label{eq:pot_pl_ob1}~~~.
\end{eqnarray}
On the other hand, if particles injected inside the wavy current sheet structure~($\alpha_{\rm inc} - \pi/2 \le \theta \le \pi/2$), from Eqs.~(\ref{eq:phi_w_out}) and (\ref{eq:phi_w_in}), the potential difference, $\Delta \Phi_{\rm w} = \Phi_{\rm w}(\theta) - \Phi_{\rm w}(0)$, is 
\begin{eqnarray}
\Delta \Phi_{\rm w} &=& \frac{u_{\rm sh}}{c} \frac{R_{\rm A} \Omega_*}{V_{\rm w}} B_{\rm A} R_{\rm A} \nonumber \\
&&\times \left\{ \cos \theta - \frac{2}{\pi} \cos \theta \cos^{-1}\left( \frac{\cos \theta}{\sin \alpha_{\rm inc}} \right) \right. \nonumber \\
&& \left. + \frac{2}{\pi} \sqrt{\sin^2 \alpha_{\rm inc} - \cos^2 \theta} - 1 \right\} \label{eq:pot_pl_ob2}~~~.
\end{eqnarray}
Therefore, particles injected at the equator are accelerated to the following maximum energy 
\begin{eqnarray}
\varepsilon_{\rm max,PD} &=& e \Delta \Phi_{\rm w}(\theta = \pi/2) \nonumber \\
&=& \left(\frac{2}{\pi} \sin \alpha_{\rm inc} - 1\right) \frac{u_{\rm sh}}{c} \frac{R_{\rm A} \Omega_*}{V_{\rm w}} e B_{\rm A} R_{\rm A} \label{eq:emax_pot_ob_pl}~~.
\end{eqnarray}
It should be noted that $B_{\rm A}$ is negative for $\alpha_{\rm inc} \ge \pi/2$.

\begin{figure}[h]
\centering	
\includegraphics[scale=0.7]{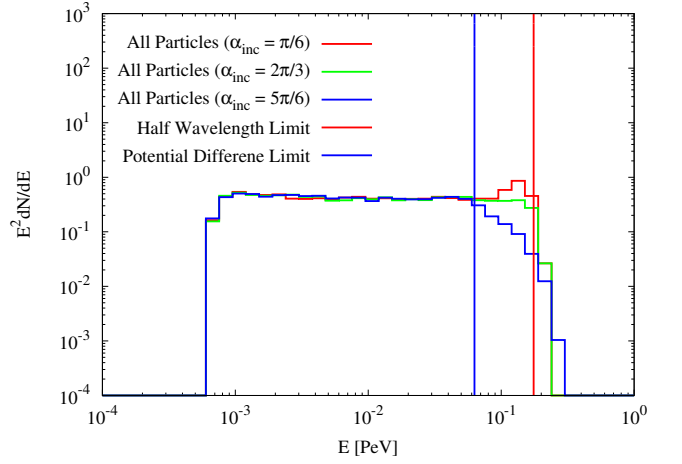}
\caption{Energy spectra for $t_{\rm inj} = 0.3~{\rm yr}$, RSGs with  $\alpha_{\rm inc} = \pi/6, 2\pi/3$, and $5\pi/6$. The spectra are generated from the simulation results at $t \approx 0.5~{\rm yr}$ when the cutoff energy is saturated. The red, green, and blue histograms are for $\alpha_{\rm inc} = \pi/6, 2\pi/3$, and $5\pi/6$, respectively.
The red and blue vertical lines show the maximum energies at $t \approx 0.5~{\rm yr}$, which are limited by the half wavelength of the wavy current sheet [Eq.~(\ref{eq:half})] and the potential difference [Eq.~(\ref{eq:emax_pot_ob_pl})], respectively.
\label{fig:es}}
\end{figure}
\begin{figure}[h]
\centering	
\includegraphics[scale=0.7]{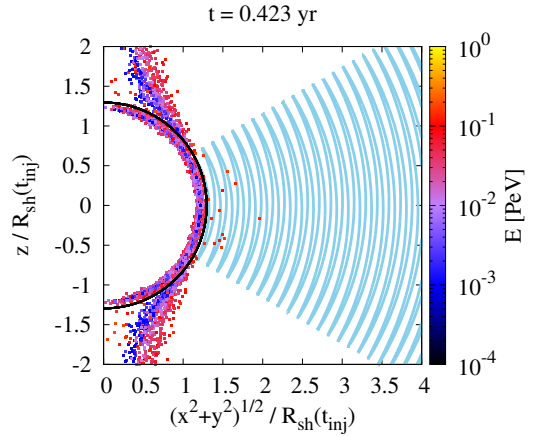}
\caption{Same as Fig.~\ref{fig:posi_rsg_pl_ob_e}, but for $\alpha_{\rm inc} = 5\pi/6$. The figure shows the distribution of particles at $t = 0.423~{\rm yr}$.
\label{fig:posi_rsg_pl_ob_e_150deg}}
\end{figure}
Figure~\ref{fig:emax_pl} shows the maximum energy for oblique rotators with $\alpha_{\rm inc} \ge \pi/2$. 
The blue dotted and green dot-dashed lines are the maximum energy limited by the whole potential difference for $\alpha_{\rm inc} =  5\pi/6$ and $2\pi/3$, respectively.
The blue open squares and green open triangles are the simulation results for $\alpha_{\rm inc} =  5\pi/6$ and $2\pi/3$, respectively.
The black dashed line is the maximum energy limited by the half wavelength of the wavy current sheet.
As one can see in Fig.~\ref{fig:emax_pl}, the maximum energy at $t_{\rm inj} = 0.3~{\rm yr}$ for $\alpha_{\rm inc} =  5\pi/6$ is limited not by the half wavelength of the wavy current sheet but by the potential difference.

Figure~\ref{fig:es} shows energy spectra of all particles for $t_{\rm inj} = 0.3~{\rm yr}$.
The spectra are generated from the simulation results at $t \approx 0.5~{\rm yr}$ when the cutoff energy is saturated.
The red, green, and blue histograms are for $\alpha_{\rm inc} = \pi/6, 2\pi/3$, and $5\pi/6$, respectively.
The red and blue vertical lines show the maximum energies limited by the half wavelength of the wavy current sheet [Eq.~(\ref{eq:half})] and the potential difference [Eq.~(\ref{eq:emax_pot_ob_pl})], respectively.
As one can see in Fig.~\ref{fig:es}, the cutoff energies of $E^2 dN/dE$ for $\alpha_{\rm inc} = \pi/6$ and $2\pi/3$ are limited by the half wavelength of the wavy current sheet.
On the other hand, for $\alpha_{\rm inc} = 5\pi/6$, the maximum value of the cutoff tail and the cutoff energy of $E^2 dN/dE$ is almost in good agreement with the maximum energy limited by the half wavelength of the wavy current sheet and the potential difference, respectively.
The reason why the spectrum for $\alpha_{\rm inc} = 2\pi/3$ is different from that for $\alpha_{\rm inc} = 5\pi/6$ is because the number of particles inside the wavy current sheet structure is different between the cases for $\alpha_{\rm inc} = 2\pi/3$ and $\alpha_{\rm inc} = 5\pi/6$.
Figure~\ref{fig:posi_rsg_pl_ob_e_150deg} shows the distribution of particles injected on the SNR surface at $t_{\rm inj} = 0.3~{\rm yr}$ for $\alpha_{\rm inc} = 5\pi/6$.
Compared with Fig.~\ref{fig:posi_rsg_pl_ob_e_150deg} and the left panel of Fig.~\ref{fig:posi_rsg_pl_ob_e}, the number of particles inside the wavy current sheet structure for $\alpha_{\rm inc} = 5\pi/6$ is smaller than that for $\alpha_{\rm inc} = 2\pi/3$.
The area of the wavy current sheet for $\alpha_{\rm inc} =  5\pi/6$ is narrower than that for $\alpha_{\rm inc} =  2\pi/3$. 
Since particles drift to the pole, the particles easily leave from this narrow wavy current sheet structure.
Therefore, almost all particles leave from the wavy current sheet structure and the maximum energy is limited by the potential difference.

From Secs.~\ref{sec:ob_eq} and \ref{sec:ob_pl}, for oblique rotators, there are two types of the maximum energy, $\varepsilon_{{\rm max}, \lambda/2}$ and $\varepsilon_{\rm max,PD}$, and the maximum energy is determined by the larger of $\varepsilon_{{\rm max}, \lambda/2}$ and $\varepsilon_{\rm max,PD}$.
As one can see in Figs.~\ref{fig:emax_eq} and \ref{fig:emax_pl}, for both RSGs and WR stars, the typical energy scale is $10-100~{\rm TeV}$.

\section{Maximum Energy Estimated From Stellar Evolution} \label{sec:stellar}

In Sec.~\ref{sec:oblique}, we estimated the maximum energy of CRs by using observed values of RSGs and WR stars.
However, the number of samples of RSGs and WR stars that we can estimate the magnetic field strength is not enough, so that observational biases are concerned.
On the other hand, O stars are well observed compared with RSGs and WR stars.
Because the mass loss due to their stellar winds is negligible for typical RSGs, 
we can assume the magnetic flux conservation as long as the stellar dynamo does not work. 
In contrast to RSGs, it is very difficult to estimate the surface magnetic field strength of WR stars because the magnetic flux is not conserved during the stellar evolution even though the stellar dynamo does not work. 
In this section, to estimate the maximum energy of CRs, we deduce the physical values of RSGs from the observational values of O stars in the main sequence phase because RSGs evolve from O stars. 
About 10\% of O stars have strong magnetic fields of the order of $10~{\rm G} - 1~{\rm kG}$ \cite{fossati15, grunhut17, schneider19}. 
On the other hand, for the other 90\% of O stars, the magnetic field strength has not been estimated.  
Therefore, we consider only O stars that the magnetic field is estimated.

The surface magnetic field of O stars is set to be $1~{\rm kG}$ in this work \cite{fossati15, grunhut17, schneider19}.
As representative values of other physical parameters of O stars that the magnetic field strength is estimated, the metallicity, stellar mass and radius, rotation velocity, and wind velocity are set to be $Z=Z_\odot, M_{\rm O}=15M_\odot, R_{\rm O}=10R_\odot, v \sin i=100~{\rm km/s}$, and $V_{\rm w,O}=2000~{\rm km/s}$, respectively \cite{fullerton06,groh13}.
The rotation period of O stars, $P_{*,{\rm O}}$, is estimated as $P_{*,{\rm O}} = 2\pi R_{*,{\rm O}}/v \sin i \approx 5~{\rm days}$.
From the magnetic flux conservation, the surface magnetic field of RSGs is 
\begin{eqnarray}
B_{*,{\rm RSG}} &=& B_{*,{\rm O}} \left( \frac{R_{*,{\rm O}}}{R_{*,{\rm RSG}}} \right)^2~~, \nonumber \\
&\approx& 1~{\rm G} \left( \frac{B_{*,{\rm O}}}{1~{\rm kG}} \right) \left( \frac{R_{*,{\rm O}}}{10R_\odot} \right)^2 \left( \frac{R_{*,{\rm RSG}}}{300R_\odot} \right)^{-2}~~. 
\label{eq:brsg_stellar}
\end{eqnarray}
The radius of RSGs is suggested as $R_{*,{\rm RSG}} \approx 100 - 1500R_\odot$ and is set to be $300R_\odot$ in this work \cite{levesque05}.
The magnetic field strength of some RSGs is observed as $B_{*,\rm RSG} \approx 1-10~{\rm G}$, which is consistent with the above estimation \cite{tessore17, mathias18, auriere10}.

Next, we consider the rotation period of RSGs.
The angular momentum of a star with the radius, $R_*$, mass, $M$, and angular velocity, $\Omega_*$, is $J = (2/5) M R_*^2 \Omega_*$.
Rotating stars lose their angular momentum by their winds. 
The loss rate is given by $(dJ/dt)_{\rm w} = (2/3) \dot{M} R_{\rm A}^2 \Omega_*$ \cite{weber67}, where $\dot{M}$ and $R_{\rm A}$ are the mass loss rate and the Alfv{\'e}n radius. 
Then, the spin-down time of O stars, $t_{\rm sd,O} = J/(dJ/dt)_{\rm w}$, is estimated to be
\begin{eqnarray}
t_{\rm sd,O} &=& \frac{3}{5} \left( \frac{R_{\rm A,O}}{R_{\rm *,O}} \right)^{-2} \frac{M_{\rm O}}{\dot{M}_{\rm O}} \nonumber \\
&\sim& 9.4 \times 10^7~{\rm yr} \left( \frac{R_{\rm A,O}}{20R_{\rm *,O}} \right)^{-2} \left( \frac{M_{\rm O}}{15M_\odot} \right) \nonumber \\
&&\times \left( \frac{\dot{M}_{\rm O}}{2.4\times10^{-10}M_\odot/{\rm yr}} \right)^{-1}~~,
\end{eqnarray}
where $R_{\rm A,0}=\eta_{\rm *,O}^{1/4}R_{\rm *,O}$ is used because of $\eta_{\rm *,O}>1$ \cite{bjorklund21}.
The lifetime of O stars is $t_{\rm life,O}\approx 1.2\times10^{7}~{\rm yr}\left( M_{\rm O}/15M_\odot \right)^{-2.5}$, which is much smaller than the spin-down time of the O star, $t_{\rm sd,O}$. 
Thus, the angular momentum of the RSG at the initial RSG phase, $J_{\rm RSG} = (3/5) M_{\rm RSG} R_{\rm *,RSG}^2 \Omega_{\rm *,RSG}$ is almost the same as that of the O star.
The rotation period of the RSG at the initial RSG phase is given by 
\begin{eqnarray}
P_{\rm *,RSG} = 12~{\rm yr} \left(\frac{P_{\rm *,O}}{5~{\rm days}}\right) \left( \frac{R_{\rm *,RSG}}{300R_\odot} \right)^2 \left( \frac{R_{\rm *,O}}{10R_\odot} \right)^{-2}~~~, 
\end{eqnarray}
where $M_{\rm RSG} = M_{\rm O}$ is used because the mass loss during the lifetime of the O star is negligible.
For the case of $M_{\rm RSG} = 15M_\odot, \dot{M}_{\rm RSG} = 10^{-5}M_\odot/{\rm yr}, B_{\rm *,RSG} = 1~{\rm G}, R_{\rm *,RSG} = 300R_\odot$ and $V_{\rm w,RSG} = 10~{\rm km/s}$, 
the Alfv\'en point is very close to the stellar surface, $R_{\rm A,RSG}\approx R_{\rm *,RSG} $ ($\eta_{\rm *,RSG}\approx 1$).
Then, the spin-down time of the RSG is estimated to be $t_{\rm sd,RSG}=9.0 \times 10^5~{\rm yr}$.
The lifetime of the RSG, $t_{\rm life,RSG} \approx 10^5~{\rm yr}$ \cite{davies17, mauron11}, is shorter than the spin-down time of the RSG, $t_{\rm sd,RSG}$.
Therefore, the rotation period of RSGs, $P_{\rm *,RSG} = 12~{\rm yr}$, is almost constant in the whole RSG phase.

The estimated values for RSGs are $B_{*,\rm RSG} = 1~{\rm G}, R_{*,\rm RSG} = 300R_\odot, \dot{M}_{\rm RSG} = 10^{-5}M_\odot/{\rm yr}, P_{*,\rm RSG} = 12~{\rm yr}$, and $V_{\rm w, RSG} = 10~{\rm km/s}$.
In addition, to determine dynamics of SNRs, the explosion energy and ejecta mass are assumed to be $E_{\rm SN} = 10^{51}~{\rm  erg}$ and $M_{\rm ej} = 13M_\odot$.
From Eq.~(\ref{eq:half}), the maximum energy limited by the half wavelength of the wavy current sheet is 
%
\begin{eqnarray}
\varepsilon_{{\rm max}, \lambda/2} \approx 1~{\rm PeV} \left( \frac{B_{*,\rm RSG}}{1~{\rm G}} \right) \left( \frac{R_{*,\rm RSG}}{300R_\odot} \right)^2 \left( \frac{R_{\rm sh}}{10^{-4}~{\rm pc}} \right)^{-1}~~~,\label{eq:emax28}
\end{eqnarray}
%
where we used $R_{\rm A,RSG} \approx R_{*,\rm RSG}$ and $B_{\rm A,RSG} \approx B_{*,\rm RSG}$. 
The maximum energy becomes about $1~{\rm PeV}$ at $t \approx 10^{-2}~{\rm yr}$ ($R_{\rm sh}(t = 10^{-2}~{\rm yr}) \approx 10^{-4}~{\rm pc}$). 
The timescale of $10^{-2}~{\rm yr}$ is smaller than the cooling time due to the interaction between CR protons and ambient protons in the RSG wind, which is estimated as follows:
\begin{eqnarray}
t_{\rm cool,pp} &=& \frac{m_{\rm H}}{c \sigma_{\rm pp} \rho_{\rm w}(R_{\rm sh}) }~~, \nonumber \\
&=& \frac{4\pi m_{\rm H} V_{\rm w,RSG} R_{\rm sh}^2}{c \sigma_{\rm  pp} \dot{M}_{\rm RSG}}~~, \nonumber \\
&\approx& 1.1\times 10^{-1}~{\rm yr}  \left( \frac{\dot{M}_{\rm RSG}}{10^{-5}M_\odot/{\rm yr}} \right)^{-1} \nonumber \\ 
&&\times \left( \frac{V_{\rm w,RSG}}{10^6~{\rm cm/s}} \right) \left( \frac{R_{\rm sh}}{10^{-4}~{\rm pc}} \right)^2 ~~~,
\end{eqnarray}
where the cross section for the proton-proton interaction is set to be $\sigma_{\rm pp}=3\times10^{-26}~{\rm cm}^2$.
Therefore, PeV CRs can be accelerated at $t=10^{-2}~{\rm yr}$ without magnetic field amplification in the upstream region. 
The scaling given in Eq.~(\ref{eq:emax28}) is valid until $t = 0.6~{\rm yr}$ and $R_{\rm sh} (t = 0.6~{\rm yr}) \approx 3.4\times10^{-3}~{\rm pc}$. 
The maximum energy becomes about $30~{\rm TeV}$ at $t = 0.6~{\rm yr}$.
The maximum energy is limited by the potential difference after $t=0.6~{\rm yr}$.
From Eq.~(\ref{eq:emax_pd_ob}), the maximum energy limited by the potential difference is
\begin{eqnarray}
\varepsilon_{\rm max,PD} &=& 18~{\rm TeV} \left( 1- \frac{2}{\pi} \sin \alpha_{\rm inc} \right) \left( \frac{u_{\rm sh}}{0.008c} \right) \left( \frac{B_{*,\rm RSG}}{1~{\rm G}} \right) \nonumber \\
&& \times \left( \frac{R_{*,\rm RSG}}{300R_\odot} \right)^2 \left( \frac{V_{\rm w, RSG}}{10^6~{\rm cm/s}} \right)^{-1} \left( \frac{P_{*,\rm RSG}}{12~{\rm yr}} \right)^{-1} \label{eq:emax_pd_gev}~.
\end{eqnarray}
Thus, the maximum energy becomes about $10~{\rm TeV}$ at $t \approx 10^3~{\rm yr}$ ($u_{\rm sh} \approx 8\times 10^{-3}c$) at which the shock reaches the edge of the RSG wind.

Next, we consider whether or not the observed CR flux around $10~{\rm TeV}$ and $1~{\rm PeV}$ can be produced by SNRs propagating to the RSG wind. 
The CR luminosity to supply the observed CR flux is
\begin{eqnarray}
L_{\rm CR,obs} &\sim& \frac{4\pi V_{\rm Gal}}{c t_{\rm diff}(E)} \left( E^2 \frac{dF}{dE} \right)~~, \nonumber \\
&\approx& 7.3 \times10^{40}~{\rm erg~s}^{-1} \left( \frac{E}{{\rm GeV}} \right)^{-0.36} \label{eq:Lcr_obs}~~~~~,
\end{eqnarray}
where $V_{\rm Gal} \approx 5.2\times10^{68}~{\rm cm^3}$ and $(E^2dF/dE) \approx 1.6\times10^{-3}~{\rm erg~cm^{-2}~s^{-1}~sr^{-1}} (E/{\rm GeV})^{-0.7}$ are the volume of our Galaxy and observed CR energy flux, respectively. 
The diffusion time of CRs, $t_{\rm diff}(E)$, is estimated by 
\begin{eqnarray}
t_{\rm diff}(E) \sim \frac{H^2}{D(E)} \approx 4.8\times10^{15}~{\rm s}\left( \frac{E}{\rm GeV} 
\right)^{-0.34}~~,
\label{eq:tdiff}
\end{eqnarray}
where $H \simeq 7~{\rm kpc}$ and $D(E) \approx 9.7 \times 10^{28}~{\rm cm^2\ s^{-1}} (E/{\rm GeV})^{0.34}$ are the halo size of our Galaxy and diffusion coefficient of CRs, respectively.
The values of $H$ and $D$ are estimated by recent observations about the primary-to-secondary ratio of CRs and unstable CRs \cite{evoli20}. 
From Eqs.~(\ref{eq:Lcr_obs}) and (\ref{eq:tdiff}), the required CR luminosities around $10~{\rm TeV}$ and $1~{\rm PeV}$ are
\begin{eqnarray}
L_{\rm CR,obs}(10~{\rm TeV}) &\approx& 2.7\times10^{39}~{\rm erg~s}^{-1} \\ 
L_{\rm CR,obs}(1~{\rm PeV}) &\approx& 5.1\times10^{38}~{\rm erg~s}^{-1}~.
\end{eqnarray}

First, we estimate the CR luminosity around $10~{\rm TeV}$ accelerated by a SNR propagating to the RSG wind.
The kinetic energy flux per unit area of the shock propagating into the wind region is $(1/2) \rho_{\rm w} u_{\rm sh}^3$, where $\rho_{\rm w}$ is given by Eq.~(\ref{eq:rho_w}).
The area of the whole shock surface is $4\pi R_{\rm sh}^2$.
As we mentioned above, a SNR continues to produce $10~{\rm TeV}$ CRs until $t_{\rm 10{\rm TeV}} \approx 10^3~{\rm yr}$. 
We assume that a fraction, $\eta_{\rm CR}$, of the kinetic energy flux dissipated by the shock until $10^3~{\rm yr}$ converts to the energy flux of CRs.
SNRs propagating into the RSG wind are caused by mainly type II-P supernovae whose rate is $\mathcal{R}_{\rm SN,II-P}=1.2\times 10^{-2}\ {\rm yr}^{-1}$ \cite{li11,groh13}.
Therefore, the CR luminosity around $10~{\rm TeV}$ is given by 
\begin{eqnarray}
L_{\rm 10TeV,RSG} &\sim& \frac{ \mathcal{R}_{\rm SN,II-P} \eta_{\rm CR} f_{\rm O} u_{\rm sh}^3 \dot{M}_{\rm RSG} t_{\rm 10TeV} }{2 V_{\rm w,RSG}}~~,  \nonumber \\
&\approx& 1.6\times10^{38}~{\rm erg~s}^{-1} \left( \frac{f_{\rm O}}{0.03}  \right) \left( \frac{\mathcal{R}_{\rm SN,II-P}}{0.012/{\rm yr}} \right)  \nonumber \\
&& \times \left( \frac{\eta_{\rm CR}}{0.1} \right) \left( \frac{u_{\rm sh}}{0.008c} \right)^3 \left( \frac{\dot{M}_{\rm RSG}}{10^{-5}M_\odot/{\rm yr}} \right) \nonumber \\
&& \times \left( \frac{V_{\rm w,RSG}}{10^6~{\rm cm/s}} \right)^{-1}  \left( \frac{t_{\rm 10TeV}}{1000~{\rm yr}} \right)~~, \label{eq:10tev_rsg}~~~~~
\end{eqnarray}
where $f_{\rm O}$ is a fraction of rapid rotating ($P_{*,\rm O} \le 10~{\rm days}$) O stars with a strong surface magnetic field.
As we mentioned above, it is expected that about 10\% of O stars have a strong magnetic field \cite{fossati15, grunhut17, schneider19}.
About 30\% of these O stars with a strong magnetic field would rapidly rotate ($P_{*,\rm O} \le 10~{\rm days}$), so that $f_{\rm O}$ is about $0.03$ \cite{grunhut17}. 
Then, the estimated CR luminosity, $L_{\rm 10TeV,RSG}$, is smaller than the required luminosity, $L_{\rm CR,obs}(10~{\rm TeV})$. 
Therefore, if no magnetic field amplification works in RSGs, RSG winds, and SNR shocks, 
CRs accelerated by SNRs propagating to the RSG wind cannot contribute to the observed $10~{\rm TeV}$ CRs.

As with the estimation of the CR luminosity around $10~{\rm TeV}$, we estimate the CR luminosity around $1~{\rm PeV}$.
\begin{eqnarray}
L_{\rm 1PeV,RSG} &\approx& 2.5\times10^{34}~{\rm erg~s}^{-1} \left( \frac{f_{\rm O}}{0.03}  \right) \left( \frac{\mathcal{R}_{\rm SN,II-P}}{0.012/{\rm yr}} \right)  \nonumber \\
&& \times \left( \frac{\eta_{\rm CR}}{0.1} \right) \left( \frac{u_{\rm sh}}{0.02c} \right)^3 \left( \frac{\dot{M}_{\rm RSG}}{10^{-5}M_\odot/{\rm yr}} \right) \nonumber \\
&& \times \left( \frac{V_{\rm w,RSG}}{10^6~{\rm cm/s}} \right)^{-1}  \left( \frac{t_{\rm 1 PeV}}{0.01~{\rm yr}} \right)~~~~~.
\end{eqnarray}
For parameters estimated from O stars, this is much smaller than the required luminosity, $L_{\rm CR,obs}(1~{\rm PeV})$.
In order for SNRs propagating to the RSG wind to contribute to the observed PeV CRs ($L_{\rm 1PeV,RSG}\sim L_{\rm CR,obs}(1~{\rm PeV})$), 
the shock propagating into the RSG wind has to continue to produce $1~{\rm PeV}$ CRs until about $100~{\rm yr}$. 
To make the potential-limited maximum energy $1~{\rm PeV}$, 
for example, $B_{\rm *,RSG} = 10~{\rm G}, R_{\rm *,RSG} = 1000R_\odot, V_{\rm w,RSG} = 10~{\rm km/s}, \dot{M}_{\rm RSG} = 10^{-5}M_\odot/{\rm yr}$, and $P_{\rm *,RSG} = 10~{\rm yr}$ are needed, where $\eta_{\rm *,RSG}>1$ and the potential-limited maximum energy is estimated by Eq.~(\ref{eq:emax_rsg_pd_eta>1}). 
However, to produce the RSG with these parameters, the O star initially has to have a strong magnetic field of $B_{\rm *,O} = 100~{\rm kG}$ and a short rotation period of $P_{\rm *,O} = 0.3~{\rm days}$, which are far from the observed values of O stars. 
Hence, some magnetic field amplifications in the RSG, RSG wind, or SNR shocks are needed to accelerate CRs to the PeV scale.

\section{Conditions of acceleration to $10~{\rm TeV}$ and $1~{\rm PeV}$} \label{sec:condition}
Since RSGs have convection cells, it is suggested that the origin of the large magnetic field strength of Betelgeuse is the dynamo in convection cells \cite{auriere10, dorch04}. WR stars also could have a large surface magnetic field because of some dynamo processes in O stars or WR stars. 
In this section, we investigate the condition that the potential-limited maximum energy, $\varepsilon_{\rm max,PD}$, is larger than $10~{\rm TeV}$ and $1~{\rm PeV}$. 
The potential-limited maximum energy depends on the magnetic field strength and radius of the Alfv\'en point, $B_{\rm A}$ and $R_{\rm A}$ [see Eqs.(\ref{eq:emax_pd_ob}) and (\ref{eq:emax_pot_ob_pl})], which depend on the magnetic confinement parameter, $\eta_{\rm *}$ [see Eqs. (\ref{eq:ra}) and (\ref{eq:ba})]. 
Thus, there is a critical magnetic field strength, $B_{\rm cr,\eta}$, which is given by the condition of $\eta_{*}=1$.
From Eqs.~(\ref{eq:emax_pd_ob}) and (\ref{eq:emax_pot_ob_pl}), the acceleration condition is rewritten by 
\begin{eqnarray}
P_* \approx \left(\frac{2\pi e u_{\rm sh}}{c\varepsilon_{\rm max,PD}}\right) \times \left\{ \begin{array}{ll}
B_* R_*^2 V_{\rm w}^{-1} & ~(~B_* < B_{\rm cr,\eta}~) ~~\\
B_*^{\frac{1}{2}} R_*^{\frac{3}{2}} \dot{M}^{\frac{1}{4}} V_{\rm w}^{-\frac{3}{4}} & ~(~B_* > B_{\rm cr,\eta}~) \\
\end{array} \right. 
\end{eqnarray}
If the rotation period is shorter than the above value, the maximum energy is larger than $\varepsilon_{\rm max,PD}$. 
For a weak magnetic field ($B_*<B_{\rm cr,\eta}$), the spin-down time does not depend on the magnetic field strength, which are comparable to or longer than the lifetime of RSGs and WR stars. 
Therefore, the rotation period is marginally conserved for the RSG and WR star phases.
For a strong magnetic field ($B_*>B_{\rm cr,\eta}$), the spin-down time decreases with increasing the magnetic field strength. 
If the spin-down time is shorter than the lifetime of RSGs or WR stars, then they do not keep their rotation periods. 
Therefore, for a strong magnetic field ($B_*>B_{\rm cr,\eta}$), there is the other critical magnetic field strength, $B_{\rm cr,sd}$, which is given by the condition of $t_{\rm life}=t_{\rm sd}$.

\subsection{Red supergiants}
\begin{figure}[h]
\centering	
\includegraphics[scale=0.7]{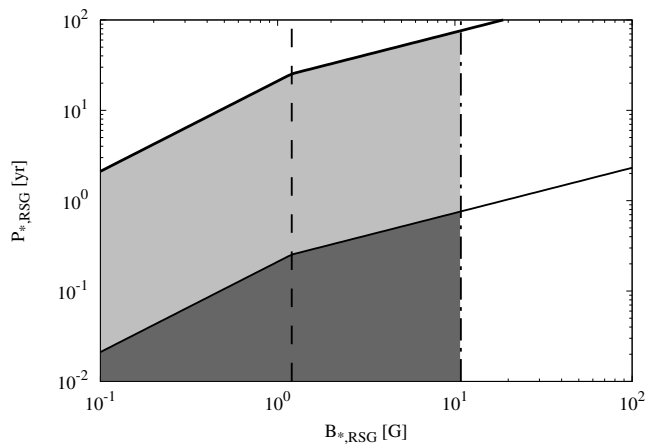}
\caption{ $P_{\rm *}-B_{\rm *}$ diagram about the acceleration condition for RSGs. 
The horizontal and vertical axes are the surface magnetic field strength and rotation period of RSGs, respectively. 
The vertical dashed and dot-dashed lines are $B_{\rm cr,\eta}$ and $B_{\rm cr,sd} $, respectively. 
The thick and thin solid lines show $\varepsilon_{\rm max,PD} = 10~{\rm TeV}$ and $1~{\rm PeV}$, respectively. 
$R_{\rm *,RSG} =  300R_\odot, M_{\rm RSG} = 15M_\odot, \dot{M}_{\rm RSG} = 10^{-5}M_\odot/{\rm yr}, V_{\rm w,RSG} = 10^6~{\rm cm/s}, u_{\rm sh} = 0.008c, t_{\rm life,RSG} = 10^5~{\rm yr}$ are used in this figure. 
If the rotation period is shorter than thick (thin) solid line, the maximum energy is larger than $10~{\rm TeV}$ ($1~{\rm PeV}$).
\label{fig:P-B_RSG}}
\end{figure}

If most RSGs have a large magnetic field strength like Betelgeuse, $f_{\rm O} = 1$ and the CR luminosity supplied in the RSG wind region is $L_{\rm RSG} \approx 5.3\times10^{39}~{\rm erg\ s^{-1}}$ [substituting $f_{\rm O} = 1$ into Eq.~(\ref{eq:10tev_rsg})].
This is comparable to the required value of $L_{\rm CR,obs}(10~{\rm TeV})$.
Thus, CRs accelerated by SNRs propagating to the RSG wind can significantly contribute to the observed CRs above $10~{\rm TeV}$ 
if the dynamo in convection cells amplifies the magnetic field of RSGs. 

For RSGs, the two critical magnetic field strength are  
\begin{eqnarray}
B_{\rm cr,\eta} &\approx& 1.2~{\rm G} \left( \frac{R_{\rm *,RSG}}{300R_\odot} \right)^{-1} \left( \frac{\dot{M}_{\rm RSG}}{10^{-5}M_\odot/{\rm yr}} \right)^{\frac{1}{2}} \nonumber \\
&&\times \left( \frac{V_{\rm w,RSG}}{10^6~{\rm cm/s}}\right)^{\frac{1}{2}} \label{eq:b_rsg_crit1}~~, \\
B_{\rm cr,sd} &\approx& 11~{\rm G} \left( \frac{t_{\rm life,RSG}}{10^5~{\rm yr}} \right)^{-1} \left( \frac{R_{\rm *,RSG}}{300R_\odot} \right)^{-1} \left( \frac{M_{\rm RSG}}{15M_\odot} \right) \nonumber \\
&&\times \left( \frac{\dot{M}_{\rm RSG}}{10^{-5}M_\odot/{\rm yr}} \right)^{-\frac{1}{2}} \left( \frac{V_{\rm w,RSG}}{10^6~{\rm cm/s}} \right)^{\frac{1}{2}} ~~.\label{eq:b_rsg_crit2}
\end{eqnarray}
Figure~\ref{fig:P-B_RSG} shows the $P_{\rm *,RSG}-B_{\rm *,RSG}$ diagram that shows acceleration conditions for $\varepsilon_{\rm max,PD} \ge 10~{\rm TeV}$ and $\varepsilon_{\rm max,PD} \ge 1~{\rm PeV}$. 
The horizontal and vertical axes are $B_{\rm *,RSG}$ and $P_{\rm *,RSG}$, respectively.
The vertical dashed and dot-dashed lines are $B_{\rm cr,\eta}$ and $B_{\rm cr,sd} $, respectively. 
The thick and thin solid lines show $\varepsilon_{\rm max,PD} = 10~{\rm TeV}$ and $1~{\rm PeV}$, respectively. 
If the rotation period is shorter than thick (thin) solid line, the maximum energy is larger than $10~{\rm TeV}$ ($1~{\rm PeV}$).
In Fig.~\ref{fig:P-B_RSG}, we use $R_{\rm *,RSG} =  300R_\odot, M_{\rm RSG} = 15M_\odot, \dot{M}_{\rm RSG} = 10^{-5}M_\odot/{\rm yr}, V_{\rm w,RSG} = 10^6~{\rm cm/s}, u_{\rm sh} = 0.008c, t_{\rm life,RSG} = 10^5~{\rm yr}$.

Parameters estimated in Betelgeuse are $P_{\rm *,RSG}\approx 36~{\rm yr}, B_{\rm *,RSG} \approx 1~{\rm G}, R_{\rm *,RSG} \approx 760R_\odot, \dot{M}_{\rm RSG} \approx 10^{-7}M_\odot/{\rm yr}, V_{\rm w,RSG} \approx 15~{\rm km/s},  M_{\rm RSG}\approx 20M_\odot$ \cite{mauron11,auriere10,kervalla18,joyce20}, which marginally satisfy the acceleration condition for $\varepsilon_{\rm max,PD}>10~{\rm TeV}$. 
Thus, after Betelgeuse explodes at the end of its life, the SNR will accelerate CRs to $10~{\rm TeV}$ without any magnetic field amplifications.

Note that in the above argument we assumed that the spin-down time must be larger than the lifetime of RSGs. This assumption limits the magnetic field strength. However, if the magnetic field strength of RSGs becomes large just before the supernova explosion by a stellar merger or some dynamo mechanisms, PeV CRs can be accelerated by SNRs propagating to the RSG wind even though the magnetic field is not amplified around the shock. 
We need to understand the evolution of the magnetic field in RSGs and the rotation period to draw the conclusion.

\subsection{Wolf-Rayet stars}
\begin{figure}[h]
\centering	
\includegraphics[scale=0.7]{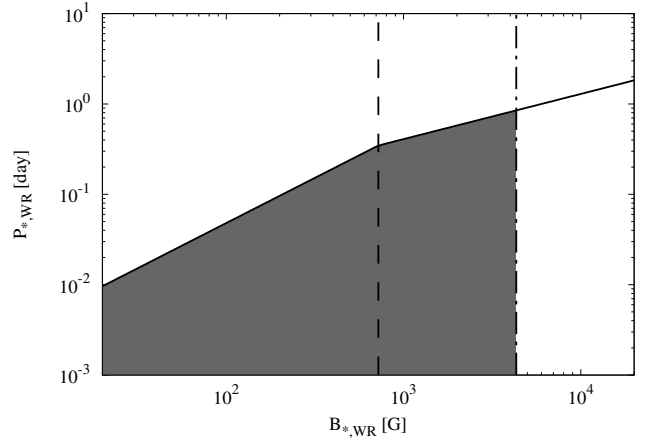}
\caption{$P_{\rm *}-B_{\rm *}$ diagram about the acceleration condition for WR stars.
The horizontal and vertical axes are the surface magnetic field strength and rotation period of WR stars, respectively.
The vertical dashed and dot-dashed lines are $B_{\rm cr,\eta}$ and $B_{\rm cr,sd} $, respectively.
The solid line show $\varepsilon_{\rm max,PD} = 1~{\rm PeV}$.
$R_{\rm *,WR} =  5R_\odot, M_{\rm WR} = 10M_\odot, \dot{M}_{\rm WR} = 10^{-5}M_\odot/{\rm yr}, V_{\rm w,WR} = 10^8~{\rm cm/s}, u_{\rm sh} = 0.018c, t_{\rm life,WR} = 10^5~{\rm yr}$ are used in this figure.
If the rotation period is shorter than the solid line, the maximum energy is larger than $1~{\rm PeV}$.
\label{fig:P-B_WR}}
\end{figure}

First, we estimate the energy dissipated by SNRs in the WR wind region.
From Eq.~(\ref{eq:emax_pd_ob}), the maximum energy limited by the potential difference, $\varepsilon_{\rm max,PD}$, is proportional to $t^{-1/8}$ for WR stars.
That is, the maximum energy is almost constant in the whole SNR age. 
In this work, we approximate that CRs with a certain energy continue to be produced until the SNR shock reaches the wind termination shock ($t \approx 1000~{\rm yr}$).
SNRs that propagate into the WR wind is caused by type Ib/Ic supernovae. 
The explosion energy of $E_{\rm ej} = 10^{51}~{\rm erg}$ and ejecta mass of $M_{\rm ej} = 8M_\odot$ are used to determine the dynamics of SNRs.
Here, it is assumed that a WR star with $10M_\odot$ loses $1M_\odot$ by the wind with the constant mass loss rate, $\dot{M}_{\rm WR} = 10^{-5}M_\odot/{\rm yr}$, during the lifetime of $t_{\rm life,WR} = 10^5~{\rm yr}$, and a neutron star with $1M_\odot$ remains after the supernova explosion \cite{meynet11,meynet03,hamann06}. 
This assumption is almost consistent with the evolution of O stars with $35M_\odot$ \cite{dwarkadas07}.
As we estimated in Sec.~\ref{sec:stellar}, the luminosity of CRs is estimated by
\begin{eqnarray}
L_{\rm WR} &\sim& \frac{ \mathcal{R}_{\rm SN,Ib/Ic} \eta_{\rm CR} u_{\rm sh}^3 \dot{M}_{\rm WR} t }{2 V_{\rm w,WR}}~~,  \nonumber \\
&\approx& 4.5\times10^{38}~{\rm erg\ s^{-1}} \left( \frac{\mathcal{R}_{\rm SN,Ib/Ic}}{0.004/{\rm yr}} \right) \nonumber \\
&&\times \left( \frac{\eta_{\rm CR}}{0.1} \right)  \left( \frac{u_{\rm sh}}{0.018c} \right)^3 \left( \frac{\dot{M}_{\rm WR}}{10^{-5}M_\odot/{\rm yr}} \right) \nonumber \\
&&\times \left( \frac{V_{\rm w,WR}}{10^8~{\rm cm/s}} \right)^{-1} \left( \frac{t}{1000~{\rm yr}} \right) \label{eq:LWR_wind}~~~,
\end{eqnarray}
where $\mathcal{R}_{\rm SN,Ib/Ic}$ and $\eta_{\rm CR}$ are the type Ib/Ic supernova rate and fraction of energy converted to CRs from the energy dissipated by the shock until $1000~{\rm yr}$, respectively.
The SNR shock reaches the termination shock of the WR wind at about $1000~{\rm yr}$.
The radius of the termination shock (the edge of the wind), $R_{\rm TS}$, is about several parsecs \cite{dwarkadas07}.
For WR winds, the total mass that the SNR shock sweeps up before reaching the termination shock, $\dot{M}_{\rm WR}R_{\rm TS}/V_{\rm w,WR}$, is of the order of $0.1M_\odot$, which is much smaller than the supernova ejecta of $M_{\rm ej}=8M_{\odot}$.
The CR luminosity produced in the WR free wind, $L_{\rm WR}$, is smaller than the required luminosity of $L_{\rm CR,obs}(10~{\rm TeV})$, but the same order of magnitude of $L_{\rm CR,obs}(1~{\rm PeV}$).
Therefore, CRs accelerated in the free wind of WR stars can contribute to the observed CRs above $1~{\rm PeV}$.

For WR stars, the two critical magnetic field strength are  
\begin{eqnarray}
B_{\rm cr,\eta} &\approx& 0.72~{\rm kG} \left( \frac{R_{\rm *,WR}}{5R_\odot} \right)^{-1} \left( \frac{\dot{M}_{\rm WR}}{10^{-5}M_\odot/{\rm yr}} \right)^{\frac{1}{2}} \nonumber \\
&&\times \left( \frac{V_{\rm w,WR}}{10^8~{\rm cm/s}}\right)^{\frac{1}{2}} \label{eq:b_wr_crit1}~~, \\
B_{\rm cr,sd} &\approx& 4.3~{\rm kG} \left( \frac{t_{\rm life,WR}}{10^5~{\rm yr}} \right)^{-1} \left( \frac{R_{\rm *,WR}}{5R_\odot} \right)^{-1} \left( \frac{M_{\rm WR}}{10M_\odot} \right) \nonumber \\
&&\times \left( \frac{\dot{M}_{\rm WR}}{10^{-5}M_\odot/{\rm yr}} \right)^{-\frac{1}{2}} \left( \frac{V_{\rm w,WR}}{10^8~{\rm cm/s}} \right)^{\frac{1}{2}} ~~.\label{eq:b_wr_crit2}
\end{eqnarray}
Figure~\ref{fig:P-B_WR} shows the $P_{\rm *,WR}-B_{\rm *,WR}$ diagram that shows acceleration conditions for $\varepsilon_{\rm max,PD} \ge 1~{\rm PeV}$. 
The horizontal and vertical axes are $B_{\rm *,WR}$ and $P_{\rm *,WR}$, respectively.
The vertical dashed and dot-dashed lines are $B_{\rm cr,\eta}$ and $B_{\rm cr,sd} $, respectively. 
The solid line show $\varepsilon_{\rm max,PD} = 1~{\rm PeV}$. 
If the rotation period is shorter than the solid line, then the maximum energy is larger than $1~{\rm PeV}$.
In Fig.~\ref{fig:P-B_WR}, we use $R_{\rm *,WR} =  5R_\odot, M_{\rm WR} = 10M_\odot, \dot{M}_{\rm WR} = 10^{-5}M_\odot/{\rm yr}, V_{\rm w,WR} = 10^8~{\rm cm/s}, u_{\rm sh} = 0.018c, t_{\rm life,WR} = 10^5~{\rm yr}$ \cite{meynet03,hamann06,crowther07,niedzielski02}.

It is suggested that the rotation period of WR 46 is about $P_{\rm *,WR} \approx 15.5~{\rm hours}$ \cite{hubrig20}.
Furthermore, the day scale rotation period is estimated for some WR stars \cite{chene08}.
Although more WR star samples that the rotation period is estimated are required, some fractions of WR stars would have the fast rotation and could accelerate CRs to the PeV scale without any magnetic field amplification around the SNR shock.
As in the case of RSGs, we need to understand the evolution of the magnetic field in WR stars and the rotation period to draw the conclusion.

\section{Discussion} 
\label{sec:discussion} 
Cassiopeia A is one of young SNRs~($t\approx300~{\rm yr}$), which is a remnant of type IIb supernova \cite{krause08}. 
Furthermore, x-ray observations suggest that Cassiopeia A is now propagating to the RSG wind \cite{lee14}. 
Therefore, Cassiopeia A is the best SNR that can directly verify this work. 
VERITAS and Fermi-LAT, and MAGIC observations reported a cutoff of the gamma-ray spectrum, showing that the maximum energy of CR protons is about 23 and $35~{\rm TeV}$, respectively \cite{ahnen17}. 
This is consistent with our results (Figs.~\ref{fig:emax_eq} and \ref{fig:emax_pl}) for RSGs.
We consider the particle acceleration in forward shocks. However, these observations cannot identify the emission region which might be the reverse shocked region. 
To confirm this work, a new gamma ray observation with a higher angular resolution is required.

As we mentioned in Sec.~\ref{sec:intro}, CREAM, NUCLEON, DAMPE, and HAWC observations reported the spectral break of CR protons and helium around $10~{\rm TeV}$.
However, the origin of the spectral break around $10~{\rm TeV}$ is still unknown.
As one can see Figs.~\ref{fig:emax_eq} and \ref{fig:emax_pl}, for both RSGs and WR stars, the maximum energy is several $10~{\rm TeV}$ in the whole SNR age. 
In addition, our recent work showed that perpendicular shock regions of typical type Ia SNRs accelerate CRs up to several $10~{\rm TeV}$ without magnetic field amplification in the shock upstream region \cite{kamijima21}. 
Thus, the origin of $10~{\rm TeV}$ CRs could be SNRs in which the upstream magnetic field amplification does not work. 
To clarify the above possibility for the origin of 10 TeV CRs, we have to understand injection processes to diffusive shock acceleration at perpendicular shocks \cite{injection}.

In this work, particles injected on the shock surface are accelerated and drift to the equator or poles.
The accumulation of the accelerated particles would modified the shock structure at the equator or poles through the pressure of the accelerated particles \cite{bell08}.
Furthermore, particles escape from the shock front along the equatorial plane or poles.
The CR streaming along the equator or poles could drive magnetic field amplifications in the wind region, which would boost the maximum energy of CRs \cite{bell04}.
So far, the magnetic field amplification by streaming CRs were investigated in an uniform medium. 
However, especially for oblique rotators, the wind region is not uniform due to the wavy current sheet. 
It is quite interesting how magnetic fields are amplified and how the amplified magnetic fields change the escape process and the maximum energy, which should be investigated in future.

We have considered the escape process of CRs accelerated in the free wind region. 
After escape from the free wind region, the escaping CRs eventually interact with the shocked region of the wind termination shock, where magnetic fields are expected to be turbulent. 
If the escaping CRs are trapped in the shocked region, then the escaping CRs could be reaccelerated by the SNR shock \cite{zirakashvili18}. 
Furthermore, massive stars explode in various environment.
The binary fraction of massive stars is suggested to be about 50\% \cite{sana13}.
In addition, most massive stars are generated in star clusters \cite{higdon05}, which is also suggested by the observed CR data \cite{ohira11}.
We have to investigate CR acceleration and escape processes in those various environment in future.

This work can be applied to transrelativitic supernovae (hypernovae, broad line type Ic SNe, and so on) propagating into their wind. 
Since their shock velocities are faster and the maximum energy would be larger than those of normal supernovae, the transrelativistic supernova are expected to be the source of CRs at $E\sim 10^{18} {\rm eV}$ \cite{ran08}. 
However, as seen in Fig.~\ref{fig:P-B_WR}, WR stars have to have a quite rapid rotation to make the maximum energy $10^{18} {\rm eV}$ even though the shock velocity is transrelativistic.
Therefore, again, some magnetic field amplifications are required in order for the transrelativistic supernovae to accelerate CRs to $10^{18} {\rm eV}$.

As mentioned in Sec.~\ref{sec:condition}, the magnetic field strength and rotation period of RSGs and WR stars are very important parameters to estimate the maximum energy of CRs accelerated in the wind region of these stars. 
However, the number of RSGs and WR stars that the magnetic field strength and rotation period are estimated is small. 
Although precise observations of O stars give us some useful information, precise observations of RSGs and WR stars are strongly desired because stellar mergers and some dynamo processes would change the rotation period and magnetic field strength.

In this work, we assumed that the downstream magnetic field strength, $B_{\rm d}$, is 100 times larger than the upstream magnetic field strength, $B_{\rm w}$. 
Accelerating particles move diffusively and drift towards the pole (or equator) in the downstream and upstream regions, respectively.
Even if the magnetic field amplification in the downstream region is modest, our results for the maximum energy and escape process do not change as long as the drift timescale is shorter than the dynamical timescale of the SNR and the diffusion timescale between the pole and equator. 
The dynamical, diffusion and drift timescales are estimated by $R_{\rm sh}/u_{\rm sh}$, $\pi^2R_{\rm sh}^2/(4\kappa_{\rm d})$ and $\pi R_{\rm sh}/(2v_{\rm \theta})$, where $\kappa_{\rm d}$ is the Bohm diffusion coefficient in the downstream region and $v_\theta$ is the drift velocity of accelerating particles. 
Since the accelerating particles drift their gyroradius in the upstream region, $r_{\rm g}$, to the $\theta$ direction during a back-and-forth motion, the drift velocity is given by
\begin{eqnarray}
v_\theta = \frac{r_{\rm g}}{\Delta t_{\rm u} + \Delta t_{\rm d}} = \frac{c}{\pi} \left\{ 1 + \frac{16}{3\pi} \left( \frac{B_{\rm d}}{B_{\rm w}} \right)^{-1} \left( \frac{u_{\rm sh}}{c} \right)^{-1} \right\}^{-1}, \label{eq:vtheta}~~
\end{eqnarray}
where $\Delta t_{\rm u}=\pi r_{\rm g}/c$ and $\Delta t_{\rm d}=16 \kappa_{\rm d}/(u_{\rm sh} c)$ are residence times in the upstream and downstream regions, respectively. 
For $B_{\rm d} > 8.4 B_{\rm w}$, the drift timescale is shorter than the dynamical and diffusion timescales for typical SNRs, so that our results in this work do not change.
On the other hand, for $B_{\rm d} < 8.4 B_{\rm w}$, the dynamical timescale is shorter than the drift timescale, so that deceleration of the shock velocity is significant while accelerating particles drift to the pole or equator. Then, the maximum energy is smaller than the estimate of Eqs.~(\ref{eq:emax_pd_ob}) and (\ref{eq:emax_pot_ob_pl}). However, the escape process does not change as long as accelerating particles reach the pole or equator.

In this work, the magnetic field turbulence is assumed to be uniform and sufficiently strong in the downstream region. 
As the result, the energy spectrum of accelerated particles is the same as the standard DSA prediction \cite{kamijima20}.
However, in reality, it takes a finite time to stretch magnetic field lines by turbulence. 
Furthermore, the turbulent magnetic field decays in the far downstream region \cite{pohl05}. 
If the magnetic field fluctuation is not sufficiently large in the downstream diffusion region of accelerating particles, $4\kappa_{\rm d}/u_{\rm sh}$, 
the accelerating particle distribution is anisotropic in the diffusion region. 
If so, then the energy spectrum of accelerated particles is softer than the standard DSA prediction \cite{takamoto15,bell11}. 
Since the diffusion region depends on the particle momentum, the energy spectrum of accelerated particles could deviate from a single power-law distribution in perpendicular shocks.

The magnetic pole of RSGs and WR stars could flip as the Sun reverses its magnetic pole \cite{babcock59}.
Although we did not consider the pole flip in this work, it would change the maximum energy and escape process because the pole flip makes other current sheets.
We are going to investigate effects of the pole flip to the escape process and maximum energy in the next paper.


\section{Summary} \label{sec:summary}
In this work, by performing test particle simulations, we investigated CR acceleration and escape from perpendicular shock regions of spherical shocks in circumstellar media.
In the shock upstream region (free wind region), the magnetic field was assumed to be the Parker-spiral magnetic field with a current sheet. 
No magnetic field amplification was considered in the upstream region, but strong turbulent magnetic fields were assumed in the shock downstream region.  
If the angle between the rotation axis and magnetic axis of progenitors is smaller (larger) than 90 degrees, accelerating particles drift to the equator (pole).
Our simulation results (Secs.~\ref{sec:aligned} and \ref{sec:oblique}) are summarized as follows:
\begin{enumerate}
\item For cases that accelerating particles drift to the equator (pole), particles escape to the far upstream along the equatorial plane (pole).
\item If progenitors are aligned rotators, the escape-limited maximum energy is given by the potential difference between the equator and pole, which is $10-100~{\rm TeV}$ for RSGs and WR stars with the magnetic field strength and rotation period expected from observations of RSGs and WR stars. 
\item If progenitors are oblique rotators, there is a wavy current sheet structure in the free wind region. In the early phase of SNRs ($t\sim 1~{\rm yr}$ for RSGs and $t\sim 10^{-2}~{\rm yr}$ for WR stars), the escape-limited maximum energy can be larger than the potential difference because of the wavy structure, which is limited by the condition that the gyroradius is the half wavelength of the wavy structure. In the later phase of SNRs, the escape-limited maximum energy is given by the potential difference between the equator and pole. 
\end{enumerate}

In Secs.~\ref{sec:aligned} and \ref{sec:oblique}, we investigated the maximum energy by using the magnetic field strength and rotation period expected from observations of RSGs and WR stars.
In Sec.~\ref{sec:stellar}, we estimated the magnetic field strength and rotation period by using  observed values of O stars, where any dynamo and spin-up processes were not considered. The estimated values of RSGs are almost in good agreement with observed values of some RSGs. 
However, if no magnetic field amplification works in RSGs, RSG winds, and SNR shocks, the CR luminosity produced in the RSG free wind is smaller than that required from CR observations. 
This is because only a small fraction of RSGs have a sufficient magnetic field strength to accelerate CRs to $10~{\rm TeV}$ and $1~{\rm PeV}$.

Since the number of observed samples of RSGs and WR stars is small, we have not understand the evolution of magnetic fields and rotation period of massive stars. 
Some dynamo processes could amplify the magnetic field strength. 
Stellar merger and accretion processes could change the rotation period. 
In Sec.~\ref{sec:condition}, instead of estimating the magnetic field strength and rotation period, we investigated conditions that the potential-limited maximum energy is larger than 10 TeV and 1 PeV.  
In addition, we estimated luminosities of CRs accelerated by SNRs propagating to the RSG and WR wind. 
Our results are shown as follows: 
\begin{enumerate}
\item The CR luminosity estimated from the dissipated energy in the RSG wind region can contribute the observed CR luminosity above $10~{\rm TeV}$ if a strong magnetic field is maintained in most RSGs.
Thus, SNRs that progenitors are RSGs could be the origin of the $10~{\rm TeV}$ break in the spectrum of CR protons and helium. 
If the surface magnetic field of RSGs is amplified just before the supernova explosion, the SNR shock propagating into RSG winds could accelerate CRs to PeV without any upstream magnetic field amplifications. 

\item For WR stars, the luminosity of CRs accelerated in the WR wind region is smaller than the required CR luminosity around $10~{\rm TeV}$, but comparable to that above $1~{\rm PeV}$. Although the rapid rotation is required for the production of PeV CRs, the rapid rotation period is actually estimated for some WR stars.
\end{enumerate}
Precise observations of RSGs and WR stars are strongly desired to measure the rotation period and surface magnetic field, which are crucial parameters to limit the maximum energy of CRs.


\begin{acknowledgments}
We thank M. Hoshino and T. Amano for valuable comments. 
We thank K. Ioka and Atoms visiting program at Yukawa Institute of Theoretical Physics for providing the research environment for this work.
Numerical computations were carried out on Cray XC50 at Center for Computational Astrophysics, National Astronomical Observatory of Japan. 
S.K. is supported by International Graduate Program for Excellence in Earth-Space Science (IGPEES), The University of Tokyo. 
Y.O. is supported by JSPS KAKENHI Grants No. JP19H01893 and No. JP21H04487.
\end{acknowledgments}

\end{document}